\documentclass[12pt,a4paper]{article}

\usepackage{epsfig}
\usepackage{amssymb}
\usepackage[centerlast,footnotesize]{caption2}
\usepackage{float}

\newcommand{\lapprox}{\raisebox{-0.5ex}{$\
\stackrel{\textstyle<}{\textstyle\sim}\ $}}
\newcommand{\gapprox}{\raisebox{-0.5ex}{$\
\stackrel{\textstyle>}{\textstyle\sim}\ $}}

\usepackage{graphicx}
\newcommand{\One}{1\kern-4.5pt1}

\newcommand{\be}{\begin{equation}}
\newcommand{\ee}{\end{equation}}

\def\lesim{${\lower 2pt\hbox{$\scriptstyle
<$}\atop\raise 4pt\hbox{$\scriptstyle\sim$}}$} 
\def\grsim{${\lower2pt\hbox{$\scriptstyle >$} \atop\raise4pt\hbox 
{$\scriptstyle\sim$}}$} 

\begin{document}
\begin{center}
\begin{flushright}
October 2016
\end{flushright}
\vskip 10mm
{\LARGE
Towards Critical Physics in 2+1$d$ with U(2$N$)-Invariant Fermions 
}
\vskip 0.3 cm
{\bf Simon Hands}
\vskip 0.3 cm
{\em Department of Physics, College of Science, Swansea University,\\
Singleton Park, Swansea SA2 8PP, United Kingdom.}
\end{center}

\noindent
{\bf Abstract:} 
Interacting theories of $N$ relativistic fermion flavors in reducible spinor representations 
in 2+1 spacetime dimensions are formulated on a lattice using domain wall
fermions (DWF), for which a U($2N$) global symmetry is recovered in the limit that the
wall separation $L_s$ is made large. The Gross-Neveu (GN) model is studied in the
large-$N$ limit and an exponential acceleration of convergence to the
large-$L_s$ limit is demonstrated if the usual parity-invariant mass
$m\bar\psi\psi$
is replaced by the U($2N$)-equivalent $im_3\bar\psi\gamma_3\psi$. The GN model 
and two lattice variants of the Thirring model are simulated for $N=2$ using a hybrid
Monte Carlo algorithm, and studies made of the symmetry-breaking bilinear
condensate and its
associated susceptibility, the axial Ward identity, and the mass spectrum of
both fermion and meson excitations. Comparisons are made with existing results
obtained using staggered fermions. 
For the GN model a symmetry-breaking phase
transition is observed, 
the Ward identity is recovered, and the spectrum found to be consistent with
large-$N$ expectations. There appears to be no obstruction to the study of
critical UV fixed-point physics using DWF. For the Thirring model the Ward identity is not
recovered, the spectroscopy measurements are inconclusive, and no symmetry
breaking is observed all the way up to the effective strong coupling limit.
This is consistent with a critical Thirring flavor number $N_c<2$, contradicting earlier 
staggered fermion results. 
\vspace{0.5cm}


\section{Introduction}
\label{sec:intro}
Relativistic field theories of particles moving in the plane have received
recent attention, particularly within the condensed-matter community, because of
their potential role in describing the spin-liquid phase of quantum
antiferromagnets~\cite{Wen:2002zz}, the pseudogap phase of cuprate
superconductors~\cite{Herbut:2002yq}, and of course low-energy electronic
excitations in
graphene~\cite{CNGPNG}.  However, they are interesting to study in their own
right. Self-interacting theories of fermions are thought to exhibit an
unexpectedly rich
variety of ultra-violet renormalisation group fixed
points~\cite{Gehring:2015vja}, each yielding a new
interacting continuum theory. One manifestation of the fixed points is the
existence of phase transitions separating massless fermions from a phase in
which a mass gap is dynamically generated; in condensed matter physics this
represents a transition from a metallic to an insulating phase. Different 
fixed points fall in different universality classes, which depend
on both the pattern of symmetry breaking and the number of interacting species
$N$.

Because the fixed points occur at strong coupling, they present a calculational
challenge,
exemplified by the
generic power-counting non-renormalisability of the perturbative expansion in powers of $g^2$
for spacetime dimensionality $d>2$, since for a four-fermi contact interaction
$[g^2]=2-d$. At this stage it helps to be more concrete by discussing specific
examples. 
The Gross-Neveu (GN) model for interacting fermions in 2+1d is defined by the
continuum Lagrangian density
\begin{equation}
{\cal L}_{GN}=\bar\psi(\partial\!\!\!/\,+m)\psi-{g^2\over 2N}(\bar\psi\psi)^2,
\label{eq:LGN}
\end{equation}
where $\psi$ is an $N$-flavor 4 component spinor field. The bare mass and
interaction terms each reduce the global symmetry from U($2N$) to
U($N)\otimes$U($N$) (see discussion below (\ref{eq:LThir})); in addition there is a discrete Z$_2$
symmetry\footnote{Strictly a (Z$_2)^2$ symmetry if $\gamma_3$ is taken into
account.}
$\psi\mapsto\gamma_5\psi$, $\bar\psi\mapsto-\bar\psi\gamma_5$ which is broken by
the mass term but not the interaction. Whilst a weak-coupling expansion makes no
sense as stated in the previous paragraph, it is possible to develop an
alternative expansion in powers of $1/N$, favouring Feynman diagrams containing
closed loops, and suggesting a resummation~\cite{Rosenstein:1990nm}. 
At strong coupling $ag^2\geq ag_c^2\sim O(1)$ in the limit $m\to0$,
where $a$ is a UV regulator length scale,
it is found that Z$_2$ symmetry
is spontaneously broken by a vacuum bilinear condensate
$\langle\bar\psi\psi\rangle\not=0$. There is apparently no obstruction to taking
a continuum limit $a\to0$ as $g^2\to g_c^2$ from either phase. In the resummed theory the
interaction is no longer pointlike but rather mediated by exchange of a scalar
degree of freedom propagating as $k^{-1}$ in the deep Euclidean region
$k\to\infty$; this anomalous scaling cannot correspond to a term in a local
Lagrangian. This persists at higher order in $1/N$. Critical exponents
receive $O(1/N)$ corrections, but always consistent with hyperscaling, a
consequence of $1/N$-renormalisability~\cite{Rosenstein:1990nm, Hands:1992be}.
The picture suggested by the $1/N$ expansion is confirmed by numerical
simulations, which observe the symmetry-breaking transition and extract critical
exponents for $N$ as small as
2~\cite{Karkkainen:1993ef}--\cite{Otsuka:2015iba},
or even 1 if a honeycomb lattice is used~\cite{Li:2014aoa}.

Another model of interest is the Thirring model,
in which the interaction is a contact between conserved fermion currents,
defined by the continuum
Lagrangian density
\begin{equation}
{\cal L}_{Thir}=\bar\psi(\partial\!\!\!/\,+m)\psi+{g^2\over
2N}(\bar\psi\gamma_\mu\psi)^2,\;\;\;\mu=0,1,2.
\label{eq:LThir}
\end{equation}
The Thirring model has the same global symmetries as $N$-flavor QED$_3$. The
Lagrangian (\ref{eq:LThir}) is invariant under a U($2N$) generated by
matrices rotating the $N$ flavors among themselves tensored with the 4 Dirac
matrices $\{\One,\gamma_3,\gamma_5,i\gamma_3\gamma_5\}$. The parity-invariant mass term
$m\bar\psi\psi$ is not invariant under $\gamma_3$ or $\gamma_5$ rotations, so
there is an explicit breaking U($2N)\to$U($N)\otimes$U($N$). Goldstone's theorem
implies the spontaneous breaking of this symmetry results in $2N^2$ massless
bosons, whereas there are none for the Z$_2$ GN model of the previous paragraph.
However, like the GN model the Thirring model has a renormalisable $1/N$
expansion~\cite{Parisi:1975im,Hands:1994kb}, this time with a resummed vector
mediating interactions between conserved currents with UV behaviour $\propto
k^{-1}$. The resummation is not associated with a phase transition and the
expansion can be developed for any $g^2$, implying the coupling is marginal. 
As $g^2$ is raised the mass $M_v$ associated with the small-$k$ behaviour of the
vector propagator varies from $2m$ at weak coupling to $M_v^2\sim
O(m^{4-d}/g^2)$
at strong coupling~\cite{Hands:1994kb}. As $g^2\to\infty$ this suggests
(\ref{eq:LThir}) is a theory of conserved currents interacting via massless
vector exchange, in many respects similar to QED$_3$.

However, this may not be the end of the story. Dynamical mass generation through spontaneous 
symmetry breaking does not occur to any order in $1/N$, but 
for sufficiently large $g^2$ and sufficiently small $N$
there are
self-consistent solutions of Schwinger-Dyson equations which do have this property 
~\cite{Gomes:1990ed}--\cite{Sugiura:1996xk}. In the limit $g^2\to\infty$
there is a critical $N_c$ below which symmetry breaking occurs: for integer $N<N_c$ we therefore expect to find fixed
points for some finite $g_c^2(N)$. The problem has also been studied using
the functional renormalisation group~\cite{Gehring:2015vja,Gies:2010st},
implying the existence of at least
two distinct fixed points in the space of possible fermionic theories; however,
this approach suggests the nature of the fixed-point interaction is more general
than the simple ``GN'' or ``Thirring'' forms (\ref{eq:LGN},\ref{eq:LThir}) discussed so far, and that a
more faithful description requires extra interaction terms consistent with the
global symmetries in play.  The Thirring model  has also been studied using
lattice simulations for $2\leq
N\leq18$~\cite{DelDebbio:1997dv}--\cite{Christofi:2007ye}
which confirm that a symmetry-broken phase is indeed present, that
$N_c\approx7$~\cite{Christofi:2007ye}, and that critical exponents extracted
from the equation of state close to the fixed point depend sensitively on
$N$, quite distinct from the behaviour of the GN model. Since none of these
properties is accessed in a systematic weak-coupling method, the 2+1$d$
Thirring model may well be the simplest fermionic QFT {\it requiring\/} a computational
approach.

Almost all lattice feld theory studies of 2+1$d$ fermions to date have employed
the staggered fermion formulation, in which fields are described by $N_s$-flavor 
single spinor component Grassmann fields $\chi,\bar\chi$ defined at each site, and
relativistic covariance in the long-wavelength limit ensured by including a
space-dependent $\pm$ sign on each link with the defining property that the
product of such factors around any elementary plaquette equals -1. Well-known
algebraic transformations show that a conventional Dirac action is recovered as
$a\to0$ expressed in terms of reducible (ie. 4-spinor) fields $\psi,\bar\psi$,
with $\psi$ defined not at a site but rather distributed over the $2^3$ sites
defining an elementary cube. Hence $\psi$ is interpreted as describing
$2N_s$ flavors of 4-component spinor~\cite{Burden:1986by}. However, for $a\not=0$
the staggered formulation does not respect the expected continuum U($4N_s$) symmetry but
rather a remnant U($N_s)\otimes$U($N_s$). Within the lattice community it is widely
believed that the full global symmetry is recovered in the weakly-coupled $a\to0$ limit expected
for QCD: there is no reason to believe this is also the case at a
strongly-coupled fixed point. For instance, with $N=2$ the above considerations
suggest distinct breaking patterns of Z$_2$ for the GN model and
U(4)$\to$U(2)$\otimes$U(2) for Thirring. Recent simulations performed with $N_s=1$  staggered
fermions using an efficient fermion bag algorithm, however, have found compatible
critical exponents for both ``GN''~\cite{Chandrasekharan:2013aya}  and
``Thirring''~\cite{Chandrasekharan:2011mn} lattice models, suggesting that for
this minimal flavor number the two models describe the same fixed point; in
other words, extra microscopic
interactions forced by the lower symmetry of the
staggered action~\cite{DelDebbio:1997dv}
may be pushing both models into the same renormalisation group
basin of attraction. This seems a surprising result when the models are
formulated using bosonic auxiliary fields as in the following section, which is
both natural for developing the $1/N$ expansion and required for a conventional
hybrid Monte Carlo (HMC) algorithm; however when written purely
in terms of $\chi,\bar\chi$ fields distributed over the
vertices of elementary cubes, the interactions differ by only one, presumably irrelevant,
term~\cite{Chandrasekharan:2013aya}. 

The considerations of the previous paragraph suggest staggered fermions are not
adequate to capture faithfully the correct physics of a fixed point with
$g_c^2\not=0$. An approach in which the fermions have the correct global U($2N$)
symmetry built in is strongly indicated. This insight has been shared by the
Jena group, who have recently applied a non-local SLAC derivative operator to
the Thirring model~\cite{Schmidt:2015fps}. In this paper we will apply domain
wall fermions (DWF), originally devised for the study of light quarks in
QCD~\cite{Kaplan:1992bt,Furman:1994ky}, and initially studied for 2+1$d$ systems
in \cite{Hands:2015qha}. The key idea is that a fictitious dimension $x_3$ is
introduced along which fermion propagation is governed by the operator
$\bar\psi\partial_3\gamma_3\psi$. The third dimension has a finite extent $L_s$,
with open boundaries called {\it domain walls\/} at each end labelled $\pm$. The
only terms coupling the walls are either proportional to the current mass $m$ or 
are interactions of the GN form (\ref{eq:LGN}). Under generic conditions there
are exponentially-localised zero-mode solutions of the 2+1+1$d$ Dirac equation
at each wall which are eigenstates of $P_\pm={1\over2}(1\pm\gamma_3)$. It is
thus plausible that 2+1$d$ operators and Green functions constructed from
2+1+1$d$ fields living on the walls retain the properties of a theory which is
invariant under rotations of the form $e^{i\alpha\gamma_3}$. On a lattice, if
the kinetic operator is chosen with a Wilson mass $M$ of opposite sign to $m$,
then the doubler modes generically plaguing lattice fermion formulations do not
couple to normalisable modes and are hence irrelevant~\cite{Kaplan:1992bt}. 
Moreover, in the
large-$L_s$ limit it has been shown both numerically~\cite{Hands:2015qha} and
analytically~\cite{Hands:2015dyp} that full U($2N$) symmetry is recovered,
ie. $e^{i\beta\gamma_5}$ and $e^{-\delta\gamma_3\gamma_5}$ rotations also become
invariances. An unanticipated bonus is that the approach to the large-$L_s$
U($2N$)-invariant limit is accelerated if instead of the hermitian $m\bar\psi\psi$ the
physically equivalent, but antihermitian, mass term $im_3\bar\psi\gamma_3\psi$ is
used. This playoff between the different forms of parity-invariant mass term
available for reducible spinor representations in 2+1$d$ has also recently been
exploited in a lattice study of non-compact QED$_3$~\cite{Karthik:2015sgq}.

The remainder of the paper is organised as follows. In
Sec.~\ref{sec:formulation} we review the DWF formulation for 2+1$d$ reducible
fermions, setting out the different possible parity-invariant mass terms and
the approach of the corresponding bilinear expectation values to the large-$L_s$ limit
first studied in the context of quenched QED$_3$ in \cite{Hands:2015qha}. 
Lattice versions of the GN and Thirring models using DWF are then proposed, and their
simulation using an HMC algorithm outlined. While the lattice transcription of
the GN model is fairly straightforward, based on a bosonic scalar auxiliary field
confined to the walls~\cite{Vranas:1999nx}, there are (at least) two possible ways to treat the Thirring
model, one in which a vector auxiliary field $A_\mu$ is confined to the walls,
and one
in which $A_\mu$ is defined uniformly throughout the bulk $0\leq
x_3\leq L_s$ in analogy with the treatment of gluon degrees of freedom in QCD
with DWF. 

Next, Sec.~\ref{sec:1/N} 
examines the GN gap equation in the large-$N$
limit, which predicts a fixed point and spontaneous dynamical mass generation
for $g^2>g_c^2$. We
build on the pioneering work of
Ref.~\cite{Vranas:1999nx}  by generalising their solution to the case of the
antihermitian mass term $im_3\bar\psi\gamma_3\psi$ and demonstrating an
exponential improvement in convergence to the large-$L_s$ limit as a result.
The gap equation also serves as a check to simulations of the GN model with
$N=2$ 
presented in Sec.~\ref{sec:GN}, where results for the dynamically-generated gap
$\Sigma(g^2)$ are used to monitor the approach to the large-$L_s$ limit, and
equivalence of the hermitian and antihermitian mass terms is demonstrated. The
essential question of whether symmetry breaking and critical behaviour can be probed using DWF
is also addressed through studies of the scalar susceptibility peaking in the
vicnity of the critical point, recovery of the axial Ward identity (using a
variant of the GN model with a U(1) axial symmetry), and finally for the first
time the fermion propagator obtained with DWF is used in an exploratory study of
both fermion and meson masses. While no attempt is made at a full-blown
characterisation of the nature of criticality, the results of this section
support the physical scenario outlined above and captured in the
large-$N$ expansion~\cite{Hands:1992be}. 

In Sec.~\ref{sec:Thirring} we turn
attention to the Thirring model with $N=2$, presenting results of HMC simulations of both
surface and bulk models. The symmetry-breaking bilinear condensate  is
calculated as a function of $g^2$ and evidence presented that physical couplings all
the way up to the strong-coupling limit are probed. Each model shows evidence
for the influence of interactions, as does the auxiliary boson action, but the
results differ in qualitatively important ways. Significantly, the axial Ward
identity is not respected, signalling that the relation of lattice fields and
parameters to the putative continuum theory is not yet under control, and that
at this stage it is not yet possible to state whether surface or bulk
approaches is optimal. Fermion spectroscopy in this case is hindered by large phase
fluctuations, whereas meson spectroscopy requires a much larger temporal extent
than the $L_t=24$ studied here. A robust 
finding, however, is that there is no evidence for spontaneous symmetry
breaking, ie. our results support $\lim_{m\to0}\langle\bar\psi\psi\rangle=0$,
on volumes and using lattice parameters where symmetry breaking is clearly
observed using staggered fermions~\cite{DelDebbio:1997dv}. This strongly
suggests that for a theory of the form (\ref{eq:LThir}), $N_c<2$. The results
are summarised and discussed in Sec.~\ref{sec:discussion}, and an Appendix
contains technical details of the free fermion propagator using DWF, needed for
the large-$N$ calculation of Sec.~\ref{sec:1/N}.

\section{Formulation and Simulation}
\label{sec:formulation}
First, let's define the lattice action to be studied.
The fermion kinetic term uses the
$2+1d$ domain wall operator defined in~\cite{Hands:2015qha, Hands:2015dyp}:
\begin{equation}
S_{\rm kin}=\bar\Psi D\Psi\equiv
\sum_{x,y}\sum_{s,s^\prime}\bar\Psi(x,s)[\delta_{s,s^\prime}D_W(x\vert
y)+\delta_{x,y}D_3(s\vert
s^\prime)]\Psi(y,s^\prime)+m_iS_i,
\label{eq:SDWF}
\end{equation}
where the
fields $\Psi,\bar\Psi$ are four-component spinors defined in 2+1+1
dimensions.
The first term $D_W$ is the $2+1d$ Wilson operator defined on spacetime volume
$V$
\begin{equation}
D_W(M)_{x,y}=-{1\over2}\sum_{\mu=0,1,2}
\left[(1-\gamma_\mu)U_\mu(x)\delta_{x+\hat\mu,y}+(1+\gamma_\mu)U^\dagger_\mu(y)\delta_{x-\hat\mu,y}
\right]
+(3-M)\delta_{x,y},
\label{eq:Ddw}
\end{equation}
with $M$ the domain wall height parameter,
and $D_3$ controls hopping along the third dimension separating the domain walls at
$s=1$ and $s=L_s$: 
\begin{equation}
D_{3\,s,s^\prime}
=-\left[P_-\delta_{s+1,s^\prime}
(1-\delta_{s^\prime,L_s})
+P_+\delta_{s-1,s^\prime}(1-\delta_{s^\prime,1})\right]
+\delta_{s,s^\prime}.
\label{eq:D3dw}
\end{equation}
Here the projectors $P_\pm\equiv{1\over2}(1\pm\gamma_3)$ and the connection
link fields $U_\mu$ will be specified more fully later, with $U_\mu\equiv1$ for free
fields.
The mass term in (\ref{eq:SDWF}) only involves fields on the domain walls
themselves, and can be chosen as a linear combination of terms which are either hermitian:
\begin{equation}
m_hS_h=m_h\sum_x\bar\Psi(x,L_s)P_-\Psi(x,1)+\bar\Psi(x,1)P_+\Psi(x,L_s),
\label{eq:mhSh}
\end{equation}
or antihermitian:
\begin{eqnarray}
m_3S_3&=&im_3\sum_x\bar\Psi(x,L_s)\gamma_3P_-\Psi(x,1)+\bar\Psi(x,1)\gamma_3P_+\Psi(x,L_s);\label{eq:m3S3}\\
\mbox{or}\;\;
m_5S_5&=&im_5\sum_x\bar\Psi(x,1)\gamma_5P_-\Psi(x,1)+\bar\Psi(x,L_s)\gamma_5P_+\Psi(x,L_s).
\label{eq:m5S5}
\end{eqnarray}

Next consider 2+1$d$ fields $\psi,\bar\psi$  defined on the walls as follows:
\begin{equation}
\psi(x)=P_-\Psi(x,1)+P_+\Psi(x,L_s);\;\;\;
\bar\psi(x)=\bar\Psi(x,L_s)P_-+\bar\Psi(x,1)P_+.
\label{eq:4to3}
\end{equation}
in terms of which the three mass terms are written $m_h\bar\psi\psi$,
$im_3\bar\psi\gamma_3\psi$, and $im_5\bar\psi\gamma_5\psi$ respectively.
The three terms are all parity-invariant and define physically equivalent ways
of breaking U(2$N)\to$U($N)\otimes$U($N$). In Ref.~\cite{Hands:2015qha} it was
demonstrated, in the context of quenched QED$_3$, that for sufficiently large $L_s$ 
the three mass terms yield compatible results for the corresponding bilinear
condensates $\langle\bar\psi\Gamma_i\psi\rangle$, with
$\Gamma_i\in\{\One,i\gamma_3,i\gamma_5\}$, consistent with the recovery
of U(2$N$)-invariance in this limit. Moreover, while
$i\langle\bar\psi\gamma_3\psi\rangle$ and $i\langle\bar\psi\gamma_5\psi\rangle$
are numerically indistinguishable, the finite-$L_s$ errors show a distinct
hierarchy, ie. with
\begin{eqnarray}
\langle\bar\psi\psi\rangle_{L_s}&=&\langle\bar\psi\psi\rangle_{L_s\to\infty}+\Delta_h(L_s)+\epsilon_h(L_s);
\nonumber\\
i\langle\bar\psi\gamma_{3,5}\psi\rangle_{L_s}&=&i\langle\bar\psi\gamma_{3,5}\psi\rangle_{L_s\to\infty}+
\epsilon_{3,5}(L_s).\label{eq:residuals}
\end{eqnarray}
then $\Delta_h\gg\epsilon_h\gg\epsilon_3\equiv\epsilon_5$. The error $\Delta_h$
is defined by the imaginary part of $i\langle\bar\psi\gamma_3\psi\rangle$
obtained using just $\bar\Psi(L_s)$ and $\Psi(1)$: the 
fields on the opposite walls yield the conjugate. Therefore
$\Delta_h$ can be estimated using measurements made with mass term $m_3S_3$.
In Ref.~\cite{Hands:2015dyp} it was shown for a gauge
theory that an expansion of the bilinear condensate
$\langle\bar\psi\psi\rangle_{L_s}$
in powers of $m_h/D_{L_s}$, where $D_{L_s}$ is a 2+1$d$ {\it truncated overlap
operator} proportional to the continuum Dirac operator in the
large-$L_s$
long-wavelength limit, for finite $L_s$ generically contains even powers of $m_h$, whereas the
corresponding expansion of $i\langle\bar\psi\gamma_3\psi\rangle$ only contains
odd powers of $m_3$, a property shared with the continuum theory.
Hence a residual $\Delta_h(m_h,Ls)$ with only weak dependence on $m_h$ as
$m_h\to0$ cannot be excluded, consistent with the hierarchy reported below
(\ref{eq:residuals}).

The Gross-Neveu (GN) model for interacting fermions in 2+1$d$, defined by the
continuum Lagrangian density (\ref{eq:LGN}) 
with $\psi$ an $N$-flavor 4 component spinor field, exhibits spontaneous
breaking of Z$_2$. 
The model is readily generalised to
exhibit spontaneous breaking of a continuous symmetry, eg. U(1), by modifying the
contact interaction to $[(\bar\psi\psi)^2-(\bar\psi\gamma_5\psi)^2]$ -- see
Sec.~\ref{sec:Ward} below.
It is convenient to reformulate (\ref{eq:LGN}) in terms of a real
scalar auxiliary boson field $\sigma$:
\begin{equation}
{\cal
L}_{GN\sigma}=\bar\psi(i\partial\!\!\!/\,+m+\sigma)\psi+{N\over2g^2}\sigma^2,
\label{eq:GNsigma}
\end{equation} 
in which case symmetry breaking is signalled by
$\Sigma\equiv\langle\sigma\rangle\not=0$. Note that physically equivalent models
are obtained by replacing the contact interaction of (\ref{eq:LGN}) by the
U($2N$)-equivalent forms $-(\bar\psi\gamma_3\psi)^2$,
$-(\bar\psi\gamma_5\psi)^2$, along with masses $m_3,m_5$ multiplying the
corresponding bilinears.

Formulation of the GN model on a lattice with DWF proceeds from the observation
that the interaction with the auxiliary in (\ref{eq:GNsigma}) formally resembles
a mass term~\cite{Vranas:1999nx}. The DWF formulation follows from 
(\ref{eq:Ddw}) with $U_\mu\equiv1$, (\ref{eq:D3dw}),(\ref{eq:mhSh}) with the interaction term 
defined solely in terms of fields on the domain walls, 
\begin{equation}
S_{\rm GNint}=\sum_x\sigma(x)[\bar\Psi(x,L_s)P_-\Psi(x,1)+\bar\Psi(x,1)P_+\Psi(x,L_s)],
\label{eq:SGNint}
\end{equation}
with obvious generalisations based on (\ref{eq:m3S3},\ref{eq:m5S5}). An
interesting distinction with previous work is
that here the auxiliary field variables are simply defined on the lattice sites;
in the conventional formulation using staggered fermions $\sigma$ is defined on
the sites of the {\it dual\/} lattice~\cite{Hands:1992be}.

The other interacting theory considered in this paper is the Thirring
model (\ref{eq:LThir}).
Again, it is convenient to recast the model using a real vector auxiliary field
$A_\mu$:
\begin{equation}
{\cal L}_{Thir}=\bar\psi(\partial\!\!\!/\,+iA_\mu\gamma_\mu+m)\psi+{N\over
2g^2}A_\mu^2.
\label{eq:LThirA}
\end{equation}
In this latter form the formal §resemblance of $A_\mu$ to an
abelian gauge field is manifest, although the last term in (\ref{eq:LThirA}) spoils gauge
invariance. In the $1/N$ expansion $A_\mu$ interpolates a massive
vector boson of mass $M_v$; the ratio $M_v/m$ is governed by the coupling
strength $g^2$~\cite{Hands:1994kb}. However, symmetry breaking
U($2N)\to$U($N)\otimes$U($N$) via generation of a bilinear condensate does not
occur at any order in $1/N$.

There are several variants of lattice formulation of the Thirring model, even
when using staggered fermions~\cite{DelDebbio:1997dv}. In the so-called {\it
non-compact\/} approach the interaction between fermion bilinears and the vector
auxiliary defined on the lattice links is linear; this has the virtue that only
four-fermion terms are generated on integration over $A_\mu$, making the
connection with the continuum form (\ref{eq:LThir}) as transparent as possible.
However, as shown in \cite{DelDebbio:1997dv}, this regularisation
fails to preserve transversity of the vacuum polarisation term contributing to
the $A_\mu$-propagator (ie. $\partial_\mu\Pi_{\mu\nu}=O(a^{-1})$), leading to an
additive renormalisation of $g^2$ and consequent uncertainty in identifying the
strong-coupling limit~\cite{Christofi:2007ye}. In this paper two non-compact DWF
formulations are investigated. First, by analogy with (\ref{eq:SGNint}) we study
a {\it surface\/} formulation with $U_\mu\equiv1$ in (\ref{eq:Ddw}) 
and link fields $A_\mu(x)$ defined solely on the
walls interacting with point-split bilinears:
\begin{eqnarray}
S_{\rm surf}={i\over2}\sum_{x,\mu}A_\mu(x)[\bar\Psi(x,1)\gamma_\mu
P_-\Psi(x+\hat\mu,1)
\!\!\!&+&\!\!\!\bar\Psi(x,L_s)\gamma_\mu P_+\Psi(x+\hat\mu,L_s)]
\label{eq:Thirint}\\
+A_\mu(x-\hat\mu)[\bar\Psi(x,1)\gamma_\mu
P_-\Psi(x-\hat\mu,1)
\!\!\!&+&\!\!\!\bar\Psi(x,L_s)\gamma_\mu P_+\Psi(x-\hat\mu,L_s)].\nonumber
\end{eqnarray}
Notice in this case the interaction couples fermion fields on the same wall.
Second, we push the analogy between the vector auxiliary and an abelian gauge
field by defining a {\it bulk\/} interaction between an $s$-independent $A_\mu(x)$ 
and the vector bilinear current defined for all $s$:
\begin{equation}
S_{\rm bulk}={i\over2}\sum_{x,\mu,s}A_\mu(x)[\bar\Psi(x,s)(-1+\gamma_\mu)
\Psi(x+\hat\mu,s)]
+A_\mu(x-\hat\mu)[\bar\Psi(x,s)(1+\gamma_\mu)
\Psi(x-\hat\mu,s)].
\label{eq:Thirbulkint}
\end{equation}
This differs from (\ref{eq:Thirint}) on the walls by the presence of a
formally-irrelevant remnant of the Wilson term (corresponding to the $\pm1$s in
(\ref{eq:Thirbulkint})). 
The relation with the gauge-invariant kinetic term (\ref{eq:Ddw}) with
$U_\mu=(1+A_\mu)$ is clear. 
If we regard $A_\mu$ as a gauge field, then the
distinction between (\ref{eq:Thirint}) and (\ref{eq:Thirbulkint}) is that in the
former case the 2+1+1$d$ fields are exposed to $s$-like plaquettes carrying non-zero flux at both $s=1$
and $s=L_s$, whereas in the latter case such plaquettes carry zero flux by
construction. At strong coupling the analogy may be crude; since the effective
connection 
is $(1+A_\mu)$ rather than $e^{iA_\mu}$ the $s$-plaquettes are
not constrained by unitarity, and may still fluctuate in
magnitude if not in phase. 

After introduction of auxiliary bosons 
both GN and Thirring models with DWF can be written in the form
\begin{eqnarray}
S&=& S_{\rm kin}+S_{\rm int}+S_{\rm bos}\\
&=&\sum_{i=1}^N\sum_{x,y}\sum_{s,s^\prime}\bar\Psi^i(x,s){\cal M}(x,s\vert
y,s^\prime)\Psi^i(y,s^\prime)+S_{\rm bos},\nonumber
\end{eqnarray}
ie. bilinear in the 2+1+1$d$ fields $\Psi^i,\bar\Psi^i$, where explicit flavor
indices are shown. For the bulk Thirring model interactions are encoded within
$S_{\rm kin}$ and there is no separate $S_{\rm int}$. The interaction $S_{\rm int}$ is one of
(\ref{eq:SGNint},\ref{eq:Thirint},\ref{eq:Thirbulkint}) and the 
bosonic action $S_{\rm bos}$ is an obvious lattice
generalisation of the quadratic terms in (\ref{eq:GNsigma},\ref{eq:LThirA}).
On the assumption that ${\cal M}$, ${\cal M}^\dagger$
describe similar physics, then the HMC algorithm may be used to
simulate both models starting from the equivalent pseudofermion action
\begin{equation}
S=\sum_{j=1}^{N/2}\sum_{x,y}\sum_{s,s^\prime}\Phi^{\dagger j}(x,s)({\cal
M}^\dagger{\cal
M})^{-1}(x,s\vert
y,s^\prime)\Phi^j(y,s^\prime)+S_{\rm bos}.
\label{eq:HMC1}
\end{equation}
The requirement to have a positive definite kernel means that $N$ must be chosen
even, and hence the minimal number of flavors simulable with the HMC algorithm is
$N=2$.
However, in order to obtain the correct functional measure, in general it is
necessary to correct for the effect of unphysical bulk fermion
modes~\cite{Furman:1994ky}, so that the fermion operator coincides with a
2+1$d$ overlap operator in the large-$L_s$ limit. For U(2$N$)-invariant
fermions this is
done by including for each flavor a term $\mbox{det}(D^{-1}(m_ha=1))$ in the functional
measure~\cite{Hands:2015dyp}, which may be thought of as arising from 
integration over bosonic Pauli-Villars fields with action $\zeta^\dagger
D(1)\zeta$. It is computationally efficient to use the same pseudofermion
fields $\Phi,\Phi^\dagger$ for both fermions and Pauli-Villars fields, and the
following action results:
\begin{equation}
S=\sum_{j=1}^{N/2}\sum_{x,y}\sum_{s,s^\prime}[D^\dagger(1)\Phi^j]^\dagger(x,s)({\cal
M}^\dagger{\cal
M})^{-1}(x,s\vert
y,s^\prime)[D^\dagger(1)\Phi^j](y,s^\prime)+S_{\rm boson}.
\label{eq:HMC2}
\end{equation}
Since the GN and surface Thirring models are formulated with $U_\mu=1$ in
(\ref{eq:Ddw}), the Pauli-Villars kernel $D(1)$ has no dependence on the bosonic
variables, and so can be dropped with no dynamical impact. Hence in these cases the simpler form
(\ref{eq:HMC1}) may be used, as pointed out in \cite{Vranas:1999nx}. For
the bulk Thirring model, the action (\ref{eq:HMC2}) is simulated. For all the
results presented in this paper the domain wall height is chosen to be $Ma=1$.

\section{Insights from Large $N$}
\label{sec:1/N}
One of the main questions to be addressed in this paper is how critical physics
appears when DWF are used, and what is the resulting dependence on
additional non-physical parameters introduced in the formulation such as $L_s$.
The GN model provides a good starting point because critical behaviour is
already manifest in the large-$N$ approximation, and can be accessed
analytically. This approach was first applied using DWF in \cite{Vranas:1999nx}; the
corresponding study for staggered fermions was performed in \cite{Hands:1992be}.

We start from the continuum GN model defined in (\ref{eq:GNsigma}).
Spontaneous breaking of a Z$_2$ global symmetry is signalled by the development
of a vacuum expectation $\Sigma=\langle\sigma\rangle$, which in the large-$N$
limit is  given self-consistently by
the gap equation
\begin{equation}
{N\over g^2}\Sigma+\langle\bar\psi\psi\rangle=
{N\over g^2}\Sigma-N\mbox{tr}(\partial{\!\!\!/\,}+m+\Sigma)^{-1}=0.
\end{equation}
For DWF $\sigma$ is localised on the walls according to
(\ref{eq:SGNint}),
and the gap equation becomes
\begin{equation}
{\Sigma_h\over g^2}=\mbox{tr}[P_-(D^\dagger G)(1,L_s)+P_+(D^\dagger G)(L_s,1)];
\end{equation}
the subscript $h$ denotes that we initially focus on a hermitian interaction
term, 
with the fermion propagator given by $D^\dagger G$, and the free fermion
Green function $G(p;s,s^\prime)$ where $p$ is a 2+1$d$ momentum  is derived in Appendix~\ref{app:A}.
Throughout this section units are defined such that $a=1$.
The first term in square brackets contains the product $D^\dagger(1,s)G(s,L_s)$.
Using (\ref{eq:symmetry},\ref{eq:G+}) we find
\begin{eqnarray}
G(p;s,L_s)&=&Be^{-\alpha(L_s-s)}+(P_+A_++P_-A_-)e^{-\alpha(L_s+s-2)}\label{eq:gapprop}\\
&+&(P_+A_-+P_-A_+)e^{-\alpha(L_s-s)}
+A_m(e^{-\alpha(s-1)}+e^{-\alpha(2L_s-s-1)}),\nonumber
\end{eqnarray}
where 
\begin{equation}
2\cosh\alpha={{1+b^2+\bar p^2}\over b};\;\;\;
\bar p_\mu=\sin p_\mu;\;\;b(p)=1-M+\sum_\mu(1-\cos p_\mu).
\label{eq:defs}
\end{equation}
The coefficients $B$,
$A_\pm$, $A_{m}$ are given in
(\ref{eq:A+})--(\ref{eq:Am}) with $m_h$ replaced by $m_h+\Sigma_h$,  
while
\begin{equation}
D^\dagger(1,s)=\theta(s-1)\theta(L_s-s)[-P_+\delta_{s,2}+(b-i\bar
p{\!\!\!/\,})\delta_{s,1}+(m_h+\Sigma_h)P_-\delta_{s,L_s}].
\label{eq:Ddagger}
\end{equation}
In calculating $P_-(D^\dagger G)(1,L_s)$ and $P_+(D^\dagger G)(L_s,1)$, terms proportional to $\bar
p{\!\!\!/\,}$ can be dropped since they vanish on tracing. The resulting
gap equation is 
\begin{eqnarray}
{\Sigma_h\over g^2}&=&{4\over V}\sum_p
\biggl[(m_h+\Sigma_h)(B+A_+)+bA_m\nonumber\\
&+&e^{-\alpha(L_s-1)}[2(m_h+\Sigma_h)A_m+b(B+A_++A_-)]\nonumber\\
&+&e^{-2\alpha(L_s-1)}[(m_h+\Sigma_h)A_-+bA_m]\biggr],
\label{eq:gap}
\end{eqnarray}
where the mode sum on an $L_xL_yL_t$ lattice runs over $p_{x,y}=2\pi n_{x,y}/L_{x,y}$,
$p_0=2\pi(n_0+{1\over2})/L_t$.
Note that finite-$L_s$ corrections appear at both $O(e^{-\alpha(L_s-1)})$ and
$O(e^{-2\alpha(L_s-1)})$.
In the limit $L_s\to\infty$, we take first the massless limit $m_h\to0$ and then the
limit $\Sigma\to0$ to find the critical coupling, 
using (\ref{eq:Am}):
\begin{equation}
{1\over g^2_c}={4\over V}\sum_p \left(B+A_+-{b\over\Delta}\right).
\label{eq:betac}
\end{equation}
The summand is given, using (\ref{eq:defs}),  by 
\begin{equation}
B\left(1+{{(e^\alpha-b)}\over\Delta}\right)-{b\over\Delta}={{(e^\alpha-b)}\over{e^{2\alpha}(b-e^{-\alpha})}}
={{z^2(p)}\over{\bar p^2}},
\end{equation}
where the factor $z(p)=1-be^{-\alpha}$ was introduced in \cite{Vranas:1999nx}.
\begin{figure}[H]
    \centering
    \includegraphics[width=10.5cm]{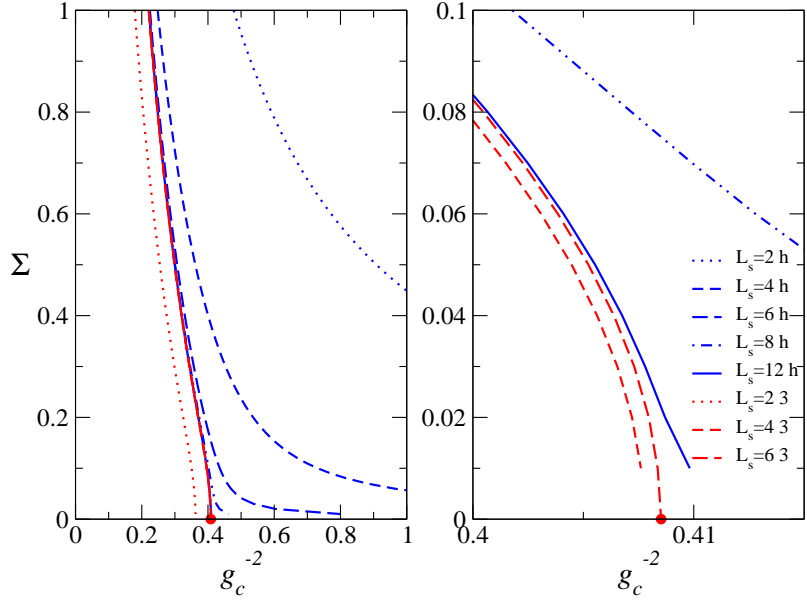}
\caption{Solution of gap equation $\Sigma(g^2)$ on a $12^3$ lattice for various
$L_s$  for both $\langle\bar\psi\psi\rangle$ ($h$) and
$i\langle\bar\psi\gamma_3\psi\rangle$ ({\it3}) (left), and the region around the critical
point enlarged
(right). In all cases $m_i=0$ and $M=1$. The red circle denotes $g_c^{-2}$ as
given by (\ref{eq:betac}).}
\label{fig:GNgap}
\end{figure}
Fig.~\ref{fig:GNgap} plots the solution to (\ref{eq:gap}) for a $12^3$
spacetime volume for various $L_s$. The red circle denotes the location of the
transition at $g_c^{-2}=0.408523$ from a massless phase to a
phase
with spontaneous mass generation given by the solution of
(\ref{eq:betac}).
It can be seen that the finite-$L_s$
corrections are significant for $L_s\lapprox6$, and still discernable even for
$L_s=12$; note that $\Sigma_h(g^2)$
approaches the large-$L_s$ limit from above. 

With mass term proportional to $m_3$ the gap equation becomes
$N\Sigma_3/2g^2-i\langle\bar\psi\gamma_3\psi\rangle=0$ and
(\ref{eq:gapprop},\ref{eq:Ddagger}) are replaced by
\begin{eqnarray}
G(p;s,L_s)&=&Be^{-\alpha(L_s-s)}+(P_+A_++P_-A_-)e^{-\alpha(L_s+s-2)}\nonumber\\
&+&(P_+A_-+P_-A_+)e^{-\alpha(L_s-s)}\\
&+&A_{m_3}e^{-\alpha(s-1)}
+A_{m_3}^*e^{-\alpha(2L_s-s-1)}\nonumber
\label{eq:gapprop3}
\end{eqnarray}
and
\begin{equation}
D^\dagger(1,s)=\theta(s-1)\theta(L_s-s)[-P_+\delta_{s,2}+(b-i\bar
p{\!\!\!/\,})\delta_{s,1}+i(m_3+\Sigma_3)P_-\delta_{s,L_s}].
\label{eq:Ddagger3}
\end{equation}
The result is 
\begin{equation}
{\Sigma_3\over g^2}=i\mbox{tr}[(P_-D^\dagger G)(1,L_s)-(P_+D^\dagger G)(L_s,1)]
\end{equation}
with
\begin{eqnarray}
(P_-D^\dagger G)(1,L_s)&=&[i(m_3+\Sigma_3)(B+A_+)+bA_{m_3}]\nonumber\\
&+&e^{-\alpha(L_s-1)}[i(m_3+\Sigma_3)(A_{m_3}+A_{m_3}^*)+b(B+A_++A_-)]\nonumber\\
&+&e^{-2\alpha(L_s-1)}[i(m_3+\Sigma_3)A_-+bA_{m_3}^*],\\
-(P_+D^\dagger G)(L_s,1)&=&[i(m_3+\Sigma_3)(B+A_+)-bA_{m_3}^*]\nonumber\\
&+&e^{-\alpha(L_s-1)}[i(m_3+\Sigma_3)(A_{m_3}+A_{m_3}^*)-b(B+A_++A_-)]\nonumber\\
&+&e^{-2\alpha(L_s-1)}[i(m_3+\Sigma_3)A_--bA_{m_3}],
\end{eqnarray}
with $A_{m_3}$ given by (\ref{eq:Am3});
the full gap equation now reads
\begin{eqnarray}
{\Sigma_3\over g^2}&=&{{4}\over V}(m_3+\Sigma_3)\sum_p
\biggl(B+A_+-{b\over\Delta_3}\label{eq:gap3}\\
&+&e^{-2\alpha(L_s-1)}\Bigl[{{2B}\over\Delta_3}\Bigl(e^{-2\alpha}(b-e^\alpha)+(m_3+\Sigma_3)^2(e^{-\alpha}-b)
+A_-\Bigr)
-{b\over\Delta_3}\Bigr]\biggr).\nonumber
\end{eqnarray}
In the large-$L_s$ limit (\ref{eq:gap3}) coincides with
(\ref{eq:gap},\ref{eq:betac}) as it must; 
remarkably, however, since $B$ is free of finite-$L_s$ corrections, and
from (\ref{eq:Delta3}) $\Delta_3$ and hence $A_\pm$ only have corrections of O($e^{-2\alpha(L_s-1)}$), 
we see that the gap equation (\ref{eq:gap3}) receives finite-$L_s$ corrections
only at this
order; use of the twisted mass in the GN system therefore  gives
exponential suppression of finite-$L_s$ corrections, as a result of
cancellation of contributions with relative phase $\pm i$ between propagators
running in opposite $s$-senses. 

\begin{figure}[thb]
    \centering
    \includegraphics[width=10.5cm]{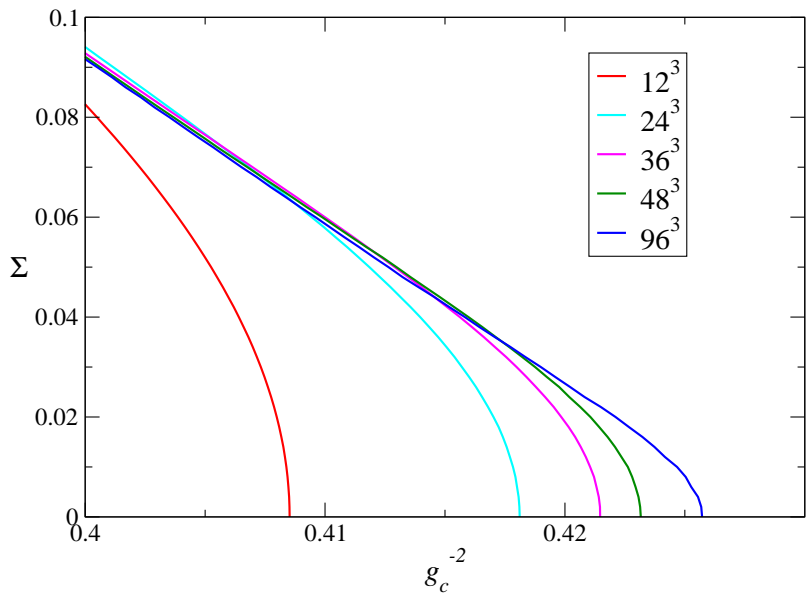}
\caption{Solution of the gap equation $\Sigma_3(g^2)$ (\ref{eq:gap3}) for various
spacetime volumes. In all cases $L_s=12$, $m_3=0$ and $M=1$.}
\label{fig:GNgapvol}
\end{figure}
Solutions of (\ref{eq:gap3}) are also plotted in
Fig.~\ref{fig:GNgap}. As predicted, the finite-$L_s$ corrections are much
smaller, and essentially under control by $L_s=6$. Also note $\Sigma_3(g^2)$
approaches the large-$L_s$ limit from below. This corroborates the improved
properties of the ``twisted mass'' formulation with respect to approaching 
the U($2N$)-symmetric limit at large $L_s$, observed empirically in quenched
QED$_3$ in \cite{Hands:2015qha}, and demonstrated analytically for gauge
theories in
\cite{Hands:2015dyp}. Finally, Fig.~\ref{fig:GNgapvol} shows the approach to the
large volume limit for fixed $L_s$; as the volume increases the expected scaling
$\Sigma_3\propto(g_c^{-2}-g^{-2})$ is recovered in the symmetry-broken phase, consistent with the large-$N$
critical exponent $\beta=(d-2)^{-1}+O(1/N^2)$~\cite{Hands:1992be}. As expected, finite volume
effects become significant for $\Sigma_3\lapprox L_x^{-1}$.

\section{Numerical Results for the Gross-Neveu Model}
\label{sec:GN}
For finite $N$ the results of the previous section are subject to 
quantum corrections.  
In principle for the GN model these are calculable via the $1/N$ expansion,
but we will use numerical simulations to address the
question of what critical behaviour of DWF fermions looks like under these
circumstances. The results 
presented in this section were obtained using a HMC
algorithm based on the action (\ref{eq:HMC1}) with the minimal
choice $N=2$, and $aM=1.0$ is used throughout. 

\subsection{Gap Equation}
\begin{figure}[thb]
    \centering
    \includegraphics[width=10.5cm]{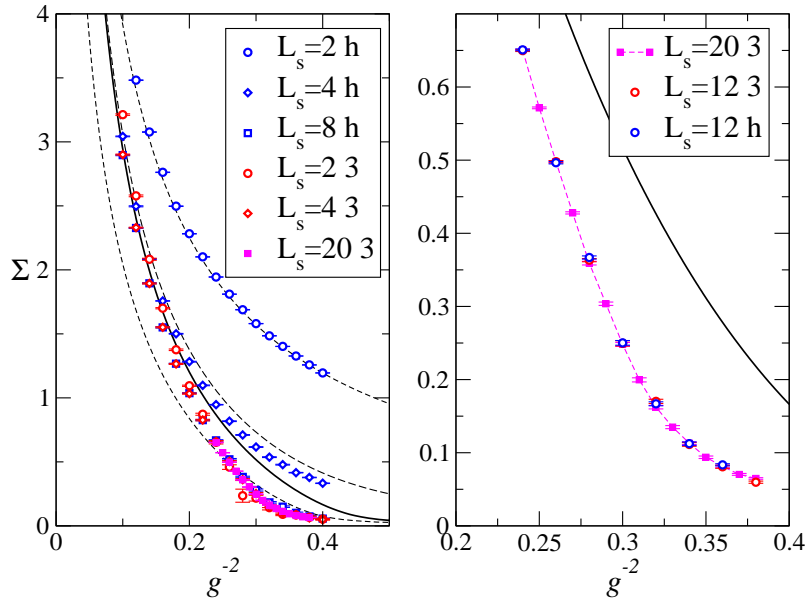}
\caption{Results for $\Sigma_{h,3}(g^2)$ obtained on a $12^3$ system for varying
$L_s$ with $am_{h,3}=0.01$. The panel at left shows the large-$N$ predictions of
(\ref{eq:gap},\ref{eq:gap3}) for $L_s=2,4$ for comparison, along with the
large-$L_s$ result as a full line. At right is data taken closer to the critical
region with larger $L_s$.}
\label{fig:gap}
\end{figure}
Initially we focus on the GN model with Z$_2$ symmetry defined by the continuum
action (\ref{eq:GNsigma}). The order parameter 
$\Sigma=\langle\sigma\rangle$ is related to the corresponding
single-flavor bilinear condensate by an equation of motion
$\Sigma_h=g^2\langle\bar\psi\psi\rangle$. The bilinear condensate may be
estimated by stochastic means~\cite{Hands:2015qha} at a significant numerical
overhead. We have checked that the simulation respects the equation of motion,
but chose to focus resources on the observable $\Sigma$.
Fig.~\ref{fig:gap} shows results for both $\Sigma_h$ and $\Sigma_3$ as functions
of $g^{-2}$ on a $12^3\times L_s$ system, with bare mass $am=0.01$. For $L_s=2$ $\Sigma_h$ is in
approximate agreement with the large-$N$ result of Section~\ref{sec:1/N}, but as
$L_s$ increases the trend is for $\Sigma(g^2)$ to fall below the
large-$N$ prediction as a result of quantum corrections. As before, the
finite-$L_s$ artifacts for $\Sigma_3$ are clearly smaller than those of $\Sigma_h$,
but in this case $\Sigma_3$ approaches the large-$L_s$ limit from above. There is fairly rapid
convergence to the large-$L_s$ limit: $\Sigma_h(L_s=8)\approx\Sigma_3(L_s=4)$.
This can be checked in closer detail in the right panel. The 
$\Sigma_3(L_s=20)$ data, shown on both panels,  were obtained from 20 - 30$\times10^3$ HMC trajectories of mean
length 1.0 and can be taken to define the effective large-$L_s$ limit. The
$L_s=12$ results for both $\Sigma_h$ and $\Sigma_3$ are consistent within statistical
errors. 

The pronounced kink in $\Sigma_3(L_s=20)$ seen in the right hand panel of
Fig.~\ref{fig:gap} hints at a critical point at $ag_c^{-2}\approx0.32$ -- 0.34,
well below the large-$N$ value predicted by (\ref{eq:betac}) as expected. Of
course, a good estimate of $g_c^{-2}$ requires demonstrable control over finite
volume artifacts and the $m\to0$ extrapolation, but such a simulation campaign
is beyond the scope of the current study. We can note, however, that finite-$N$
corrections are large, being O(50\%) in the critical region. 

\begin{figure}[thb]
    \centering
    \includegraphics[width=10.5cm]{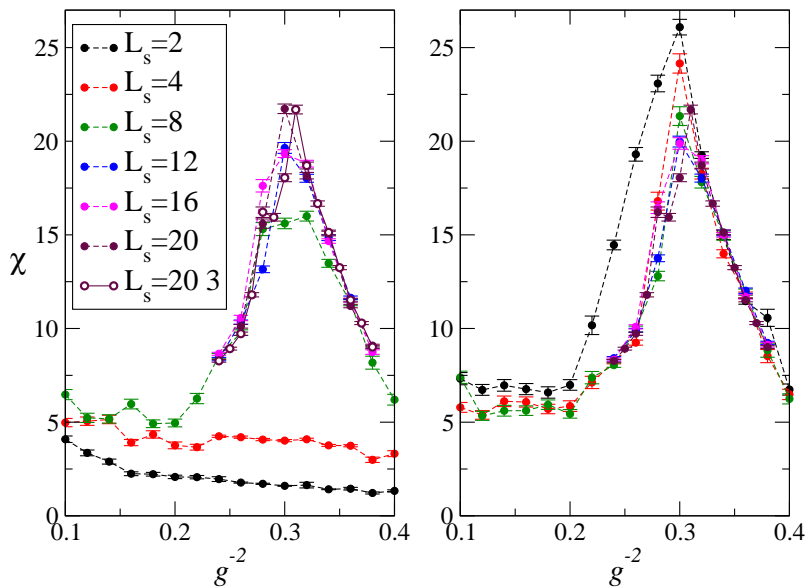}
\caption{Results for $\chi_h(g^2)$ (left) and $\chi_3(g^2)$ (right obtained on a $12^3$ system for varying
$L_s$ with $am_{h,3}=0.01$. $\chi_3(L_s=20)$ is plotted on both panels for
comparison.}
\label{fig:GNsusc}
\end{figure}

The slight increase in the size of the errorbars in the critical region just discernable in
Fig.~\ref{fig:gap} is a signal of critical fluctuations, which are more
properly quantified by the susceptibility
$\chi=\langle(\sigma-\langle\sigma\rangle)^2\rangle$, plotted in
Fig.~\ref{fig:GNsusc}. The peak in the critical region signals divergence
in the infinite volume massless limit, where we expect $\chi\propto\vert
g_c^{-2}-g^{-2}\vert^{-\gamma}$ with $\gamma=1+O(1/N)$~\cite{Hands:1992be}.
The contrast is striking: $\chi_h$
shows no sign of critical behaviour for $L_s\leq4$, and approaches the
large-$L_s$ limit from below, whereas $\chi_3$ approaches the limit from above.
The smallest value where the two are plausibly consistent is $L_s=12$, but even
for $L_s=20$ there are small differences in the data.
Unfortunately, the calculation is hard to control, particularly on the
strong-coupling side of the transition, due to occasional brief tunnelling between true and false
vacuum states related by Z$_2$ symmetry, which in this context should be
regarded as a finite volume artifact.  For this reason the actual peak height in
the large-$L_s$ limit is hard to estimate from Fig.~\ref{fig:GNsusc}. It appears
that near criticality the susceptibility presents a more stringent challenge to reaching the
large-$L_s$ limit than the order parameter.

\subsection{Axial Ward Identity}
\label{sec:Ward}
Continuous global symmetries in field theories imply the existence of Ward
identities
relating Green functions. If we wish to check restoration of a symmetry which
is formally broken at the Lagrangian level, it is important to examine the
recovery of Ward identities, to check both the symmetry itself and the
applicability of field identifications such as (\ref{eq:4to3}). To follow this
agenda in the GN model it is necessary to enhance the model by changing the
broken symmetry from Z$_2$ to U(1), requiring the introduction of a second
bosonic pseudoscalar auxiliary field $\pi$. The continuum Lagrangian becomes
\begin{equation}
{\cal
L}_{GN_{U(1)}}=\bar\psi(i\partial\!\!\!/\,+m+\sigma+i\gamma_5\pi)\psi+{N\over2g^2}(\sigma^2+\pi^2),
\label{eq:GNU1}
\end{equation} 
and the U(1) symmetry
\begin{equation}
\psi\mapsto e^{i\alpha\gamma_5}\psi;\;\;\bar\psi\mapsto\bar\psi
e^{i\alpha\gamma_5};\;\;\Phi\equiv(\sigma+i\pi)\mapsto e^{-2i\alpha}\Phi.
\end{equation}
We will focus on the axial Ward identity
\begin{equation}
{{\langle\bar\psi\psi\rangle}\over
m}=N\sum_x\langle\bar\psi\gamma_5\psi(0)\bar\psi\gamma_5\psi(x)\rangle,
\label{eq:aWard}
\end{equation}
where all Green functions are normalised to just a single fermion flavor.
In a symmetry broken phase with $\lim_{m\to0}\langle\bar\psi\psi\rangle\not=0$,
the resulting divergence of the RHS signifies the Goldstone nature of the $\pi$
field. In the GN model, the Goldstone mode is dominated 
by disconnected fermion-line diagrams 
~\cite{Barbour:1999mc}.
However, the auxiliary equations
of motion may be used to recast the identity as
\begin{equation}
\Sigma={Nm\over g^2}\chi_\pi
\label{eq:Wardbos}
\end{equation}
where $\chi_\pi$ is the transverse susceptibility
$\langle(\pi-\langle\pi\rangle)^2\rangle$, so all required expectation values
involve solely bosonic fields. Finally, note that if instead the mass term
$im_3\bar\psi\gamma_3\psi$ is
chosen then the interaction in (\ref{eq:GNU1}) takes the form
$i\bar\psi(\gamma_3\sigma-\gamma_5\pi)\psi$, but the same Ward identity
(\ref{eq:Wardbos}) results.

\begin{figure}[thb]
    \centering
    \includegraphics[width=10.5cm]{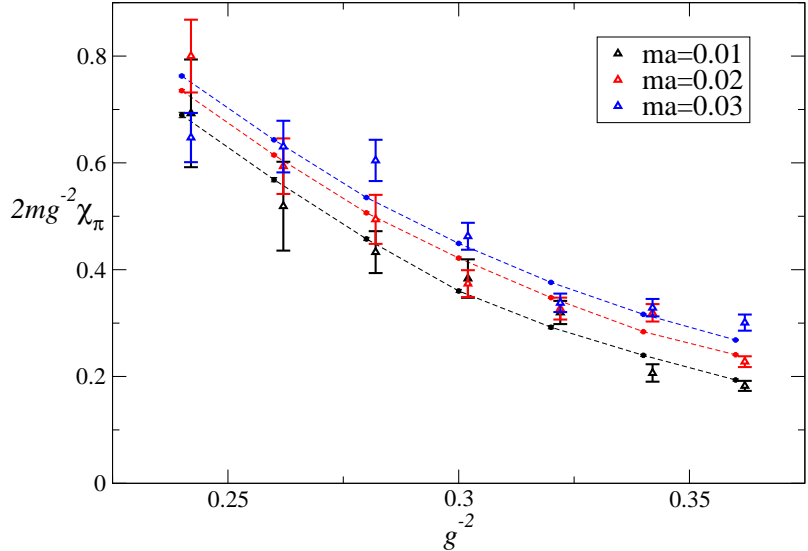}
\caption{Plot of $\Sigma$ (small symbols connected by dotted lines) and
$2m\chi_\pi/g^2$ vs. $g^{-2}$ for three values of $m$.}
\label{fig:ward}
\end{figure}
Fig.~\ref{fig:ward} shows data from the $N=2$ GN model with U(1)
symmetry and mass term $m_3S_3$ on a $12^2\times24$ spacetime lattice with $L_s=8$.
The data results from 30000 HMC trajectories of mean length 1.0.
Each side of Eqn.~(\ref{eq:Wardbos}) is plotted separately; on this scale the
errors in $\Sigma$ are hard to discern whereas the $\chi_\pi$ data suffer from
large 
fluctuations due to the Goldstone nature of $\pi$, 
similar to the staggered fermion observations of
\cite{Barbour:1999mc}. Nonetheless, within the admittedly large errors
the data are consistent with the Ward identity (\ref{eq:Wardbos}).

\subsection{Spectroscopy}
Finally we present some exploratory spectroscopy. Since the GN model
is not constrained by Elitzur's theorem, it is possible to study the propagator
of a single fermion. In addition we will examine the simplest meson correlator
formed from connected fermion lines, which as shown in ~\cite{Hands:2015qha}
interpolates states with $J^P=0^\pm$. This study uses the same
ensembles as
Sec.~\ref{sec:Ward}.

First consider the timeslice propagator of a free fermion with mass $m_f$:
\begin{equation}
\sum_{\vec
x}\langle\psi(0)\bar\psi(x)\rangle\sim\int dp_0{e^{ip_0x_0}\over{ip_0\gamma_0+m_f}}=
P_{0\pm}e^{-m_f\vert x_0\vert},
\label{eq:fermi_cont}
\end{equation}
with $P_{0\pm}\equiv{1\over2}(1\pm\gamma_0)$ and the sign chosen according to
the sign of the temporal displacement $x_0$. Using the identification
(\ref{eq:4to3}) and the identities $P_\pm P_{0+}P_\pm={1\over2}P_\pm$, 
$P_\mp P_{0+}P_{\pm}={1\over2}\gamma_0P_\pm$, the corresponding 2+1+1$d$
correlator with mass $m_h$ and $x_0>0$ is
\begin{eqnarray}
\mbox{tr}P_{0+}\biggl\langle(P_-\Psi(0,1)+P_+\Psi(0,L_s))(\bar\Psi(x,L_s)P_-\!\!\!\!&+&\!\!\!\!
\bar\Psi(x,1)P_+)\biggr\rangle=\label{eq:fermprop}
\\
{1\over2}\mbox{tr}\biggl\langle P_-\Psi(0,1)\bar\Psi(x,L_s)&+&P_+\Psi(0,L_s)\bar\Psi(x,1)
\nonumber\\
+\gamma_0\Bigl[P_-\Psi(0,1)\bar\Psi(x,1)&+&P_+\Psi(0,L_s)\bar\Psi(x,L_s)\Bigr]
\biggr\rangle.\nonumber
\end{eqnarray}
The generalisation to mass $m_3$ is straightforward.
We have measured timeslice correlators using the first and third terms of the RHS of (\ref{eq:fermprop})
using 5 randomly-located sources on configurations separated by 5 HMC
trajectories (taking care to correct for anti-periodic temporal boundary
conditions when $x_{0{\rm sink}}<x_{0{\rm source}}$). They yield two distinct
estimates of the fermion correlator labelled $\One$ (formed from 2+1+1$d$
propagators linking the two domain walls) and $\Gamma_0$ (formed from
propagators starting and ending on the same wall) in
the following. The two should coincide in magnitude if the correlator is dominated by a
simple pole of the continuum form (\ref{eq:fermi_cont}).

\begin{figure}[thb]
    \centering
    \includegraphics[width=10.5cm]{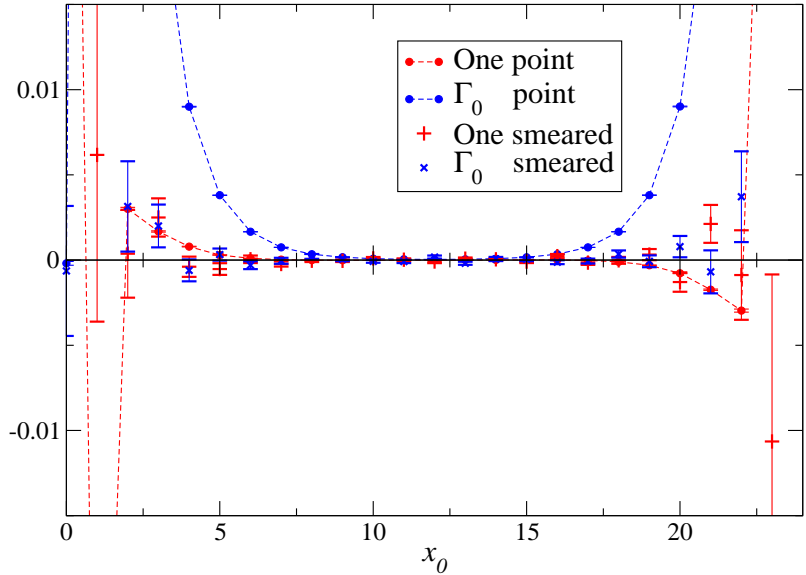}
\caption{Fermion timeslice correlators $C_{\Gamma_0}$ and $C_{\One}$  evaluated on a
$12^2\times24$ lattice with $am=0.01$ and $L_s=8$.}
\label{fig:raw}
\end{figure}
\begin{figure}[thb]
    \centering
    \includegraphics[width=10.5cm]{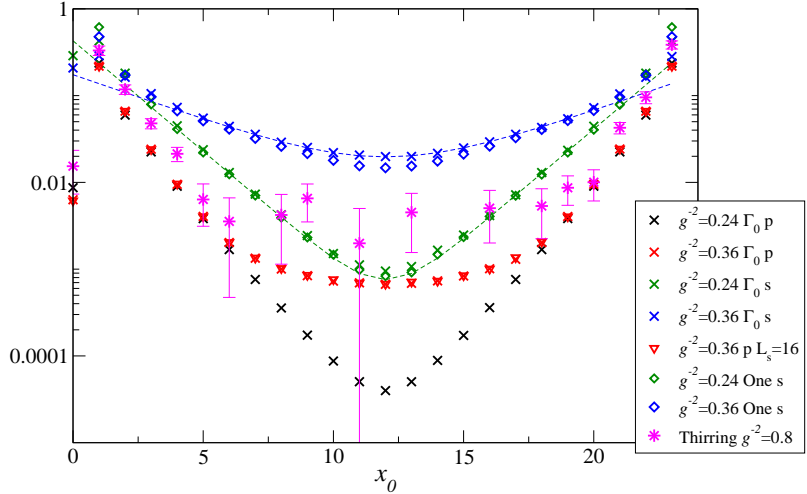}
\caption{Timeslice correlators $\tilde C_{\Gamma_0}(x_0)$ and $\tilde
C_{\One}(x_0)$ for $g^{-2}=0.24$, 0.36 and $am=0.01$.  
$L_s=8$ unless otherwise indicated. 
Results with both point ($p$) and smeared ($s$) sources are shown. Dashed lines
are
simple pole fits. Also shown are Thirring model data described in
Sec.~\ref{sec:Thirring}}
\label{fig:correlators}
\end{figure}
Fig.~\ref{fig:raw} shows the raw correlators $C_{\Gamma_0}(x_0)$ and
$C_{\One}(x_0)$
obtained at coupling $ag^{-2}=0.24$
using both a point source and a Gaussian smeared
source 
\begin{equation}
\eta^{smear}=(1-c+cD_\perp)^{N_{smear}}\eta^{point},
\label{eq:smear}
\end{equation}
with $D_\perp$ the spatial part of (\ref{eq:Ddw}). We chose $c=0.25$ and
$N_{smear}=10$. The essential feature is that $C_{\Gamma_0}$ is even
about $x_0=L_t/2$, whereas $C_{\One}$ is odd, so that their linear combination is not
symmetric about the centre of the lattice. Lines are drawn though the point
source data to emphasise this, which also suggest the two channels don't yield
signals of equivalent magnitude. The data obtained with smeared sources has the
same symmetry, but with much larger errors. These originate in the large phase
fluctuations of the background auxiliary $\Phi$ field, and render the raw
correlators useless for the precision fitting required by spectroscopy with the
available statistics. The
fluctuations also afflict the point-source data; though invisible on the scale
of Fig.~\ref{fig:raw}, the $C_{\One}$ datapoints actually have the wrong sign near the
centre of the lattice.

The pragmatic solution adopted here is instead to study the functions $\tilde
C_i=\sqrt{C^*_iC_i}$, effectively ignoring the phase fluctuations. It
should be borne in mind that $\tilde C$ thus defined is not a Green function,
and that any resulting particle mass estimate must strictly be a lower bound. 
It is also worth noting that mass fits to fermion correlators in the U(1) GN model with
staggered fermions were obtained without the need for this step~\cite{Hands:1995jq}.
Results for $\tilde C_{\Gamma_0}$ and $\tilde C_{\One}$ for two representative
couplings are plotted on a logarithmic scale in Fig.~\ref{fig:correlators}, and clearly suffer far less
from fluctuations. By construction $\tilde C$
is symmetric about the lattice midpoint.
The correlators evaluated with point sources show no coupling
dependence for $\vert x_0\vert\lapprox5$; we ascribe this to the influence at
short temporal separations of
excited states which are probably lattice artifacts. This is in notable contrast to
using staggered fermions, where excited states are absent
permitting fitting over almost the entire temporal
extent, eg.~\cite{DelDebbio:1997dv}. For this reason fits been made using smeared
sources, which yield correlators with a better projection onto the ground state
and showing a much cleaner $g^{-2}$-dependence;
reasonable fits to a simple pole were found for $x_0\in[6,18]$ ($\Gamma_0$) and $x_0\in[5,19]$
(\One), and two such fits to $\tilde C_{\Gamma_0}$ are shown. 
Most of the data of Fig.~\ref{fig:correlators} were obtained with
$L_s=8$ and mass term $m_hS_h$; we also show one correlator using $L_s=16$ and
$m_3S_3$, and it is clear that at this level of accuracy the large-$L_s$ limit
is secure. The smeared source data in Fig.~\ref{fig:correlators} also show that
$\tilde C_{\One}\approx\tilde C_{\Gamma_0}$ over the whole $x_0$ range, but that
there are systematic differences near the lattice midpoint, which may be 
a $\tilde C$-artifact; $C_{\One}$ should
vanish at the lattice midpoint, and is ideally fitted with an odd function.

\begin{figure}[thb]
    \centering
    \includegraphics[width=10.5cm]{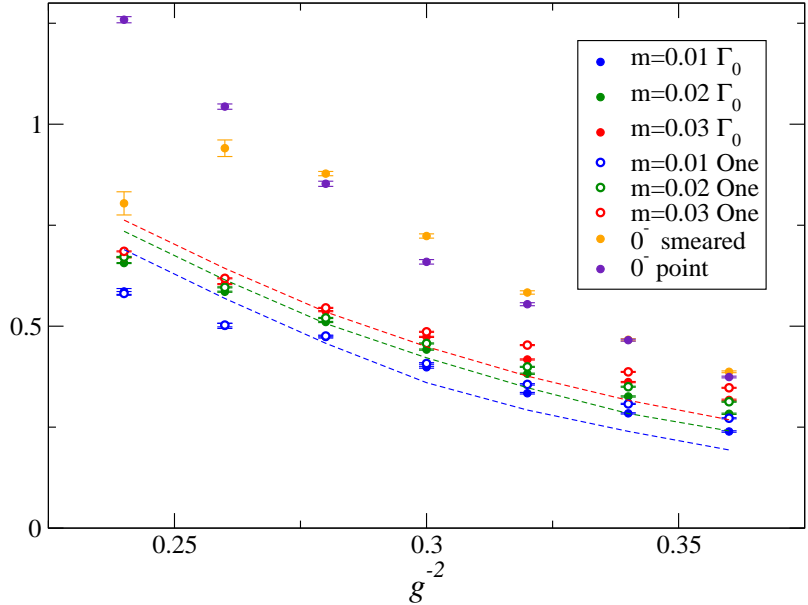}
\caption{Spectrum of the U(1) GN model as a function of $g^{-2}$, showing
results for the fermion using smeared sources for $am=0.01,.02,.03$ in both
$\Gamma_0$ and \One\/ channels, 
and the $0^-$
meson for $am=0.01$ using both point and smeared sources. Dashed lines show the $\Sigma$
data from Fig.~\ref{fig:ward}.}
\label{fig:spectrum}
\end{figure}
Fig.~\ref{fig:spectrum} shows results for the fermion mass $m_f$ for
$ag^{-2}\in[0.24,0.36]$ and $am=0.01,.02,.03$, together with the $\Sigma$ data
from Fig.~\ref{fig:ward}. In the large-$N$ limit
$m_f=\Sigma+m$~\cite{Hands:1992be}. The plot shows that this relation is approximately observed, but
the measured $m_f$ falls systematically below $\Sigma$ at strong coupling and
above $\Sigma$ at weaker coupling. The fits also yield
$m_{f\Gamma_0}<m_{f\One}$, with the trend becoming more marked at weak coupling,
as might be anticipated from Fig.~\ref{fig:correlators}. Both $m_f$ and $\Sigma$ show similar
variation with $m$ over the whole range studied. Determining whether the origin
of the mismatch is due to finite spatial volume, the fitted $x_0$-range, $O(1/N)$ corrections, 
or an artifact of 
fitting $\tilde C$ rather than the Green function $C$ is beyond the scope of this
exploratory study.

Fig.~\ref{fig:spectrum} also shows mass fits to meson correlation functions,
defined by the combination $C^{+-}+C^{--}$ using the notation of Sec. 5.2 of
Ref.~\cite{Hands:2015qha} (see (\ref{eq:C5}) 
below), corresponding to states interpolated by the
bilinears $\bar\psi\gamma_5\psi$ ($J^P=0^-$), and 
$\bar\psi\gamma_3\psi$ ($0^+$). Again, there is evidence of
significant excited state contamination (see Fig.~\ref{fig:meson_corrs} below), and the fits shown here were obtained
from $x_0\in[6,18]$. There was some difficulty in obtaining stable fits at
strong coupling using smeared sources, but by and large point and smeared
sources yield compatible
results. All the previous remarks about systematic effects apply
here; the main feature revealed in Fig.~\ref{fig:spectrum} is
$M_{0^\mp}\approx2m_f$. Although the Goldstone mode has quantum numbers $0^-$,
it is only accessed via disconnected fermion line diagrams~\cite{Barbour:1999mc}, or perhaps more
effectively via the auxiliary $\pi$ field as in Sec.~\ref{sec:Ward}. Mesons
formed from connected lines are only weakly bound by $O(1/N)$ effects,
hence the spectrum revealed in Fig.~\ref{fig:spectrum} is physically plausible.

\section{Numerical Results for the Thirring Model}
\label{sec:Thirring}
\begin{figure}[thb]
    \centering
    \includegraphics[width=10.5cm]{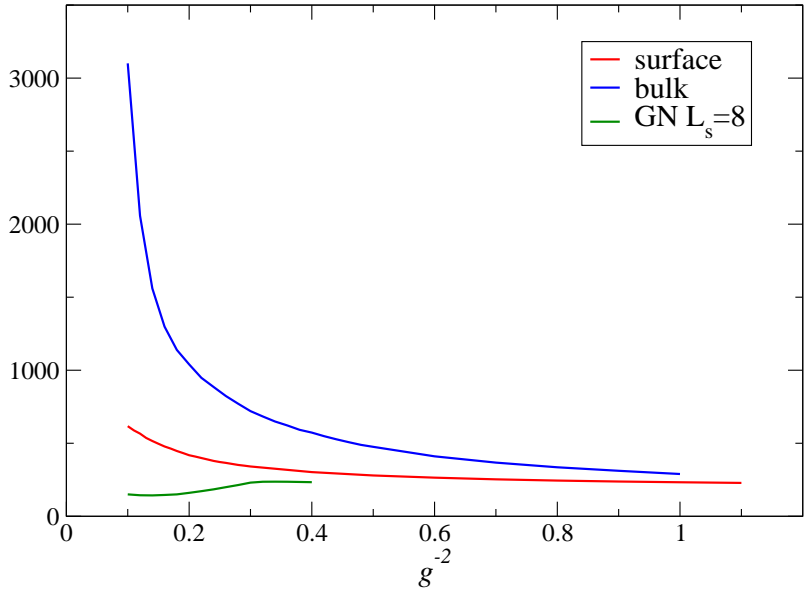}
\caption{Number of conjugate gradient iterations needed for HMC acceptance
step.}
\label{fig:congrad}
\end{figure}

Next we turn to the Thirring model, for which the bosonic auxiliary field
$A_\mu$ is not simply related to a bilinear condensate order parameter, and
where there is no straightforward analytic approach to compare with numerical
results. Moreover as discussed in Sec.~\ref{sec:formulation} the lattice
prescription is not unique. Accordingly we will explore the models defined by both surface
(\ref{eq:Thirint}) and bulk (\ref{eq:Thirbulkint}) interaction terms. Unless
otherwise stated, the results of this section were obtained with 5000 HMC
trajectories over a range of couplings $g^{-2}$ on a $12^3$ system with $L_s=16$
and $am_3=0.01$, with $N=2$ fermion flavors. The residual $\Delta_h(L_s=16)$ defined in
(\ref{eq:residuals}) ranges between 0.6 -- 1.0$\times10^{-6}$ so the results are
safely in the large-$L_s$ limit implying
$\langle\bar\psi\psi\rangle=i\langle\bar\psi\gamma_3\psi\rangle$. For reference,
Fig.~\ref{fig:congrad} plots the mean number of congugate gradient iterations
to achieve a residual norm of $10^{-9}$ per vector component,
needed in the HMC acceptance step, for each model as a function of $g^{-2}$. 
The relative cost of the bulk model rises steeply as the coupling becomes
strong. For comparison the plot also shows corresponding data for the GN model of
Sec.~\ref{sec:GN} with $L_s=8$; here by contrast the number of iterations is maximal near
the critical point.

\begin{figure}[thb]
    \centering
    \includegraphics[width=10.5cm]{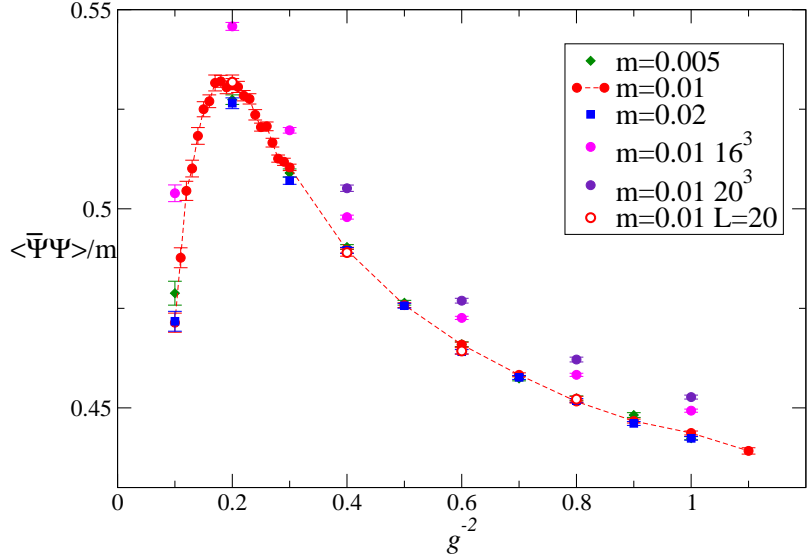}
\caption{Bilinear condensate $i\langle\bar\psi\gamma_3\psi\rangle/m_3$ vs.
$g^{-2}$ for various $am_3$, volume, and $L_s$, for the surface model
(\ref{eq:Thirint}).}
\label{fig:surface_cond}
\end{figure}
First we simulate the surface model using the pseudofermion action
(\ref{eq:HMC1}).
Fig.~\ref{fig:surface_cond} plots the ratio of the bilinear condensate
$i\langle\bar\psi\gamma_3\psi\rangle$ to the U($2N$) symmetry breaking mass
$m_3$  for values of the coupling $ag^{-2}\in[0.1,1.0]$. The expectation
value is measured using 10 stochastic estimators every 5 HMC trajectories, as
described in Sec.~5.1 of \cite{Hands:2015qha}.  Over the couplings explored the condensate
varies by about 20\%, and shows marked non-monotonic behaviour, peaking at 
$ag^{-2}\approx0.2$. The plot also includes data taken with $L_s=20$, showing
that the large-$L_s$ limit is effectively reached, and data taken on $16^3$ and
$20^3$ systems (the latter with 2000 HMC trajectories) showing significant
volume effects, though smaller than those shown in Fig.~\ref{fig:GNgapvol} for the
large-$N$ GN model in the critical region.

\begin{figure}[thb]
    \centering
    \includegraphics[width=10.5cm]{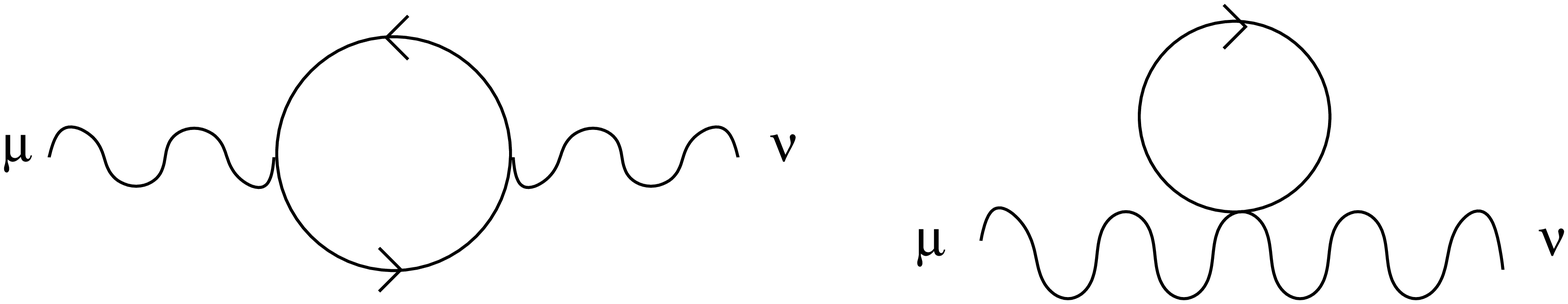}
\caption{Leading order $1/N$ contributions to the vacuum polarisation
tensor in lattice QED.}
\label{fig:vacpol}
\end{figure}

The peak in the order parameter at strong couplings has also been observed in
simulations of the Thirring model with staggered
fermions~\cite{DelDebbio:1997dv,Christofi:2007ye}. In \cite{DelDebbio:1997dv} it
was observed that the fermion-auxiliary interaction 
fails to preserve transversity of the vector current
correlator, ie. 
\begin{equation}
\sum_\mu\Pi_{\mu\nu}(x)-\Pi_{\mu\nu}(x-\hat\mu)\not=0
\end{equation}
where
$\Pi_{\mu\nu}$ is the vacuum polarisation tensor. Transversity originates in 
Ward-Takahashi (WT) identities arising from an underlying gauge symmetry, which on a
lattice implies the link field is represented by $e^{iA_\mu}$ rather than simply
$A_\mu$. In
QED$_3$ the WT identity follows from a cancellation of an $O(a^{-1})$ divergence
between the two diagrams shown in Fig.~\ref{fig:vacpol}.
With a linearised interaction of the form (\ref{eq:Thirint}) the right hand
diagram is absent because there is no 2-fermion 2-boson vertex.  The resulting
linear divergence is absorbed by an additive renormalisation of the coupling: 
$g_R^{-2}=g^{-2}-J(m,N)a^{-1}$. The physical strong coupling limit 
$g_R^{-2}\to0$ is thus found at non-zero $g^{-2}$; in practice its location
must be determined by numerical simulation~\cite{Christofi:2007ye}. For
$g_R^{-2}<0$
the vector correlator in the $1/N$ expansion becomes negative, signalling
violation of reflection positivity. 

Now, the WT identity is independent of the details
of the lattice fermion regularisation; even without a detailed calculation of the diagrams in
Fig.~\ref{fig:vacpol} using DWF it is reasonable to apply the same arguments 
to the current case. Hence we interpret the peak in Fig.~\ref{fig:surface_cond} as
evidence that the effective strong coupling limit lies at $ag^{-2}\approx0.2$, and
that the simulations have thus explored a range of couplings up to this limit.
The variation of $i\langle\bar\psi\gamma_3\psi\rangle$ with $g^{-2}$ shows clear
evidence for interaction effects.
We now observe that data taken with $am_3=0.005,0.01,0.02$ lie on top of each
other, or in other words, there is no evidence to contradict the hypothesis that
$\lim_{m_3\to0}i\langle\bar\psi\gamma_3\psi\rangle=0$ for all values of the
coupling. This is in marked contrast with results obtained using staggered
fermions on the same volume with comparable lattice parameters; compare Fig. 7
of Ref.~\cite{DelDebbio:1997dv}. We conclude that a spontaneous
symmetry breaking U($2N)\to$U($N)\otimes$U($N$) is absent in the Thirring model
defined by (\ref{eq:Thirint}) for $N=2$.

\begin{figure}[thb]
    \centering
    \includegraphics[width=10.5cm]{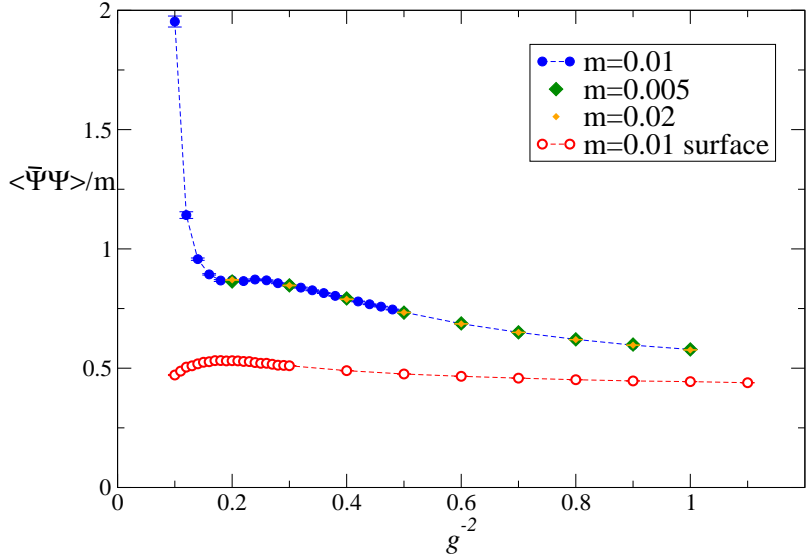}
\caption{Bilinear condensate $i\langle\bar\psi\gamma_3\psi\rangle/m_3$ vs.
$g^{-2}$ for various $am_3$ for the bulk model
(\ref{eq:Thirbulkint}). Surface model results from Fig.~\ref{fig:surface_cond}
are plotted for comparison.}
\label{fig:bulk_cond}
\end{figure}
Fig.~\ref{fig:bulk_cond} shows the results of a similar study for the bulk model
(\ref{eq:Thirbulkint}), this time using the pseudofermion action
(\ref{eq:HMC2}) to perform the HMC simulation. The magnitude of 
$i\langle\bar\psi\gamma_3\psi\rangle/m_3$ is considerably larger, reflecting the
fact that the two lattice models are different regularisations of a field
theory. Again, there is evidence for $g^{-2}$-dependence, and a local maximum
at $ag^{_2}\approx0.2$, this time followed by a steep rise at stronger
couplings. Data taken at different $m_3$ lie on top of each other
following rescaling, once again consistent with the absence of symmetry
breaking.
\begin{figure}[thb]
    \centering
    \includegraphics[width=10.5cm]{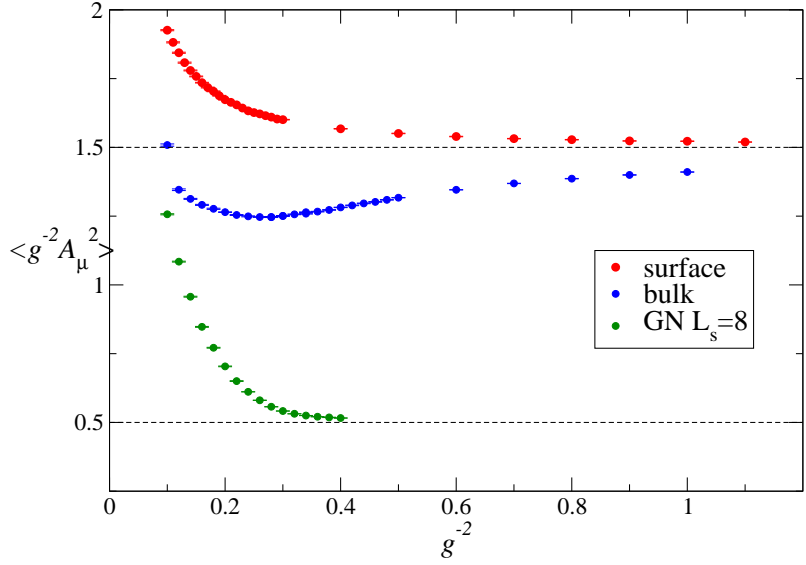}
\caption{Auxiliary boson action $g^{-2}A_\mu^2$ for both surface and bulk models
with $am_3=0.01$ on $12^3$, and for the GN model of Sec.~\ref{sec:GN} with
$L_s=8$. The dashed lines denote free-field values.}
\label{fig:boseact}
\end{figure}
An interesting contrast between the two formulations is highlighted in
Fig.~\ref{fig:boseact} plotting the boson action $g^{-2}A_\mu^2$ per lattice site.
For non-interacting fields the expected value is ${3\over2}$.
In the surface model the action density stored in the auxiliary fields exceeds
the free-field value and increases
with coupling strength, whereas the bulk model exhibits the opposite trend,
starting from the right below the free-field value and decreasing up
to the effective strong coupling limit at $ag^{-2}\gapprox0.2$. 
Large UV artifacts might be expected for the expectation value 
of a composite operator, 
and indeed this is the preferred interpretation
for what are ostensibly two different regularisations of the same theory. 
Nonetheless, the contrast between surface and bulk models may prove a useful diagnostic. For comparison the
corresponding quantity $g^{-2}\sigma^2$ is plotted for the Z$_2$ GN model of
Sec.~\ref{sec:GN}. Here there is a clear distinction between near
free-field behaviour at weak coupling and a sharp upward rise in the symmetry-broken phase,
readily understood since $\sigma$ is also an order parameter field.

Next consider the axial Ward identity as test of the extent to which U($2N$)
symmetry is restored. The equivalent identity has been found to hold in
simulations of the Thirring model with staggered
fermions~\cite{DelDebbio:1999he}.
For a $U(2N)$-invariant theory such as the Thirring model in the limit $m\to0$,
the axial Ward identity (\ref{eq:aWard}) generalises to
\begin{equation}
{{\langle\bar\psi_i\psi_i\rangle}\over
m}=\sum_{j=1}^N\sum_x\langle\bar\psi_i\gamma_3\psi_i(0)\bar\psi_j\gamma_3\psi_j(x)\rangle
=\sum_{j=1}^N\sum_x\langle\bar\psi_i\gamma_5\psi_i(0)\bar\psi_j\gamma_5\psi_j(x)\rangle
\equiv\chi_{\sigma,\pi},
\label{eq:aWard1}
\end{equation}
where no sum is implied by repeated explicit flavor indices.
The mesons interpolated by $\bar\psi\gamma_3\psi$,
$\bar\psi\gamma_5\psi$ have opposite parities.
With mass term $m_3$, the equivalent identity has $\bar\psi\gamma_5\psi$ as the
pseudoscalar, and the equivalent correlator contains contributions from 2+1+1$d$
propagators $S(m_3;0,s;x,s^\prime)=\langle\Psi(0,s)\bar\Psi(x,s^\prime)\rangle$ 
both running between the walls, and starting and ending on the same
wall~\cite{Hands:2015qha}:
\begin{eqnarray}
\langle\bar\psi\gamma_5\psi(0)\bar\psi\gamma_5\psi(x)\rangle
\equiv C_\pi^3(x)&=&{\rm
tr}\Bigl[S(m_3;0,L_s;x,L_s)P_-S^\dagger(m_3;0,L_s;x,L_s)P_+\nonumber\\
&+&S(m_3;0,1;x,1)P_+S^\dagger(m_3;0,1;x,1)P_-\nonumber\\
&+&S(m_3;0,1;x,L_s)P_-S^\dagger(m_3;0,1;x,L_s)P_-\nonumber\\
&+&S(m_3;0,L_s;x,1)P_+S^\dagger(m_3;0,L_s;x,1)P_+\Bigr]\nonumber\\
&\equiv&C^{3-+}(x)+C^{3+-}(x)+C^{3--}(x)+C^{3++}(x).\label{eq:C5}
\end{eqnarray}
Assuming that only connected fermion line diagrams
contribute to the Ward identity, we define the pion susceptibility $\chi_\pi=\sum_x C_\pi^3(x)$.

\begin{figure}[thb]
    \centering
    \includegraphics[width=10.5cm]{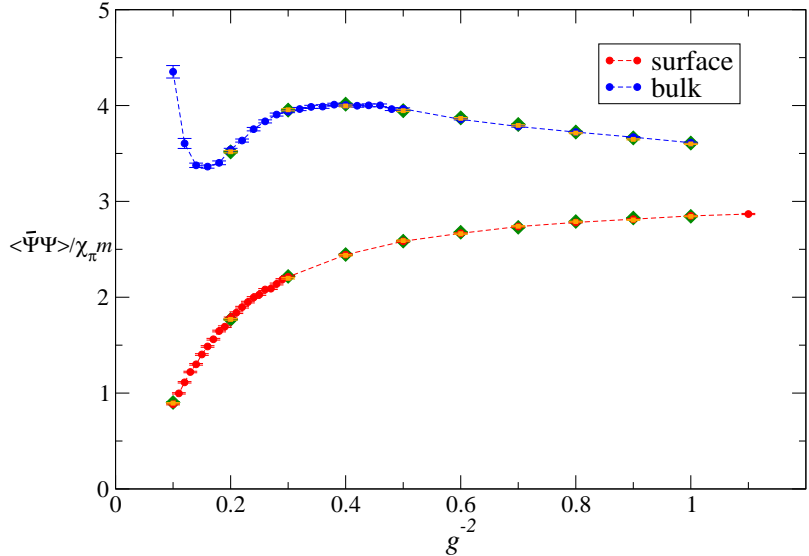}
\caption{Ratio $i\langle\bar\psi\gamma_3\psi\rangle/\chi_\pi m_3$ testing axial
Ward identity over a range of $g^{-2}$ for both surface and bulk Thirring
models, on $12^3\times16$ with $am_3=0.005,.01,.02$. Symbols are
defined as in Fig.~\ref{fig:bulk_cond}.} 
\label{fig:ratio}
\end{figure}
Fig.~\ref{fig:ratio} plots the ratio $i\langle\bar\psi\gamma_3\psi\rangle/m_3\chi_\pi$ as a
function of $g^{-2}$, which should take the value unity if the axial Ward
identity is preserved by the lattice formulation, for both surface and bulk models.
There are clear problems both in terms of the ratio's magnitude and also
its variation with $g^{-2}$, which is smooth but significant in the regime
$g_R^{-2}>0$. It is interesting that the trend is  
opposite for surface and bulk models, again suggestive that the $g^{-2}$
variation is a UV artifact. There is no variation with $m_3$.

The Ward identity is not respected because the bare action
(\ref{eq:SDWF},\ref{eq:Thirint},\ref{eq:Thirbulkint}) is not U($2N$)-invariant.
Possible causes of the breakdown could be that the
correct fermion mass in (\ref{eq:aWard1}) is not simply related to the 
lattice parameter $m_3$, or that the field identification (\ref{eq:4to3}) needs
modification, resulting in renormalisation of fermion bilinears,
once interactions are present. Whilst these are not fatal
objections, they do make it clear that care will be needed in applying DWF
techniques to this strongly-interacting system. In particular, 
Fig.~\ref{fig:ratio} provides little guidance as to whether
to choose bulk or surface formulations for further study. One possible way
forward is instead to regard the Ward identity as a relation between
renormalised quantities, so that $m$ in (\ref{eq:aWard1}) is replaced by $m_f$,
which as a spectral quantity is much better-defined. The physical fermion mass
was successfully measured in Thirring model simulations using staggered
fermions~\cite{DelDebbio:1997dv}. To this end the fermion
propagator on a 
$12^3\times24$ lattice with $L_s=16$, $am_3=0.01$ was studied using 45000 HMC
trajectories of the surface model,
with measurements made every 5 trajectories using 5 randomly chosen sources. The
best results were obtained with a smeared source (\ref{eq:smear}) with
$D_\perp$ incorporating a link connection of the form $U_\mu=e^{iA_\mu}$.
The resulting $C_{\Gamma_0}(ag^{-2}=0.8)$, where positive, is plotted in
Fig.~\ref{fig:correlators}\footnote{The corresponding $\tilde C_{\Gamma_0}(x_0)$ is
constant for $x_0\gapprox5$, showing that in contrast to GN a correct treatment of phase
fluctuations is essential to capture Thirring dynamics}. 
While there is a signal, the fluctuations are still too large to permit a
credible fit for $m_f$; this may well reflect the resemblance of the Thirring model to a
gauge theory, for which the correlator vanishes unless a gauge-fixing is
specified.

\begin{figure}[thb]
    \centering
    \includegraphics[width=10.5cm]{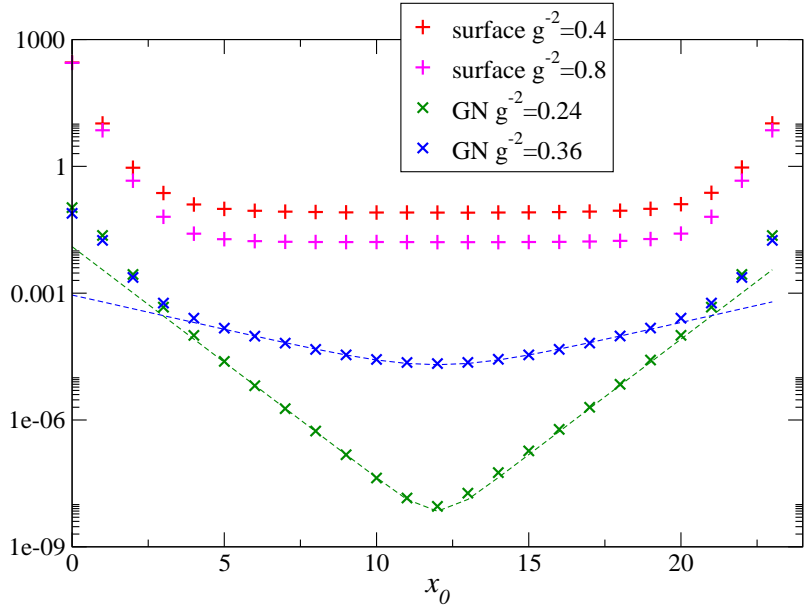}
\caption{Meson timeslice correlators $C^{+-}+C^{--}(x_0)$ obtained with point
sources 
for $g^{-2}=0.4$, 0.8 for the surface Thirring model on $12^2\times24$ with
$L_s=8$ and $am=0.01$.
Also shown are GN model data described in Sec.~\ref{sec:GN};
dashed lines
are
simple pole fits.} 
\label{fig:meson_corrs}
\end{figure}
Finally, Fig.~\ref{fig:meson_corrs} plots 
the meson correlator $C^{+-}+C^{--}$,
obtained with point sources, for the surface Thirring model
at two representative couplings. Corresponding results for the GN model
discussed in Sec.~\ref{sec:GN} yielding the spectrum plotted in
Fig.~\ref{fig:spectrum} are shown for comparison. The contast is clear; the GN
data permit a simple pole fit describing a massive meson as reported in
Fig.~\ref{fig:spectrum}, while the
$x_0$-independent plateaux in
the Thirring correlators are due to fermion propagators reconnecting after
looping around the timelike extent of the system, and are characteristic
of non-confining theories containing light fermions. Extracting spectral
information in the meson channel would require lattices of much greater temporal
extent than those studied here. Fig.~\ref{fig:meson_corrs} provides further
indirect evidence that over the coupling ranges explored dynamical fermion mass generation is happening in the GN
model, but not in the Thirring model.

\section{Discussion}
\label{sec:discussion}
Let us summarise the main results of the paper. The programme to apply DWF to
relativistic fermions in reducible spinor representations begun in
~\cite{Hands:2015qha,Hands:2015dyp} has been developed to cover non-perturbative
simulations of interesting quantum field theories. The main results of the
earlier work, namely that U($2N$) global symmetries are recovered in the
large-$L_s$ limit, and that approach to the large-$L_s$ limit is accelerated if
an antihermitian or ``twisted'' mass term $im_3\bar\psi\gamma_3\psi$ is chosen,
have been confirmed. The large-$N$ solution of the GN model presented in
Sec.~\ref{sec:1/N} provides a particularly nice illustration, 
since here the acceleration is actually exponential. Simulations
of the GN model set out in Sec.~\ref{sec:GN} provide qualitative support for
the physical picture  revealed by the large-$N$ limit, but also enable a
quantification of quantum corrections of $O(1/N)$. Crucially, the results for
the gap $\Sigma$ and scalar susceptibility $\chi$ 
demonstrate that critical physics can be observed using DWF, and that
finite-$L_s$ artifacts can be controlled. Whilst a quantitative understanding
of the critical properties and universal features of the fixed point theory
would require simulations on a much larger scale, and in particular
require much larger spacetime volumes, there is no reason to doubt the
feasibility of such a campaign.

From a theoretical perspective, the principal result of the paper is that with
$N=2$ the physics of the GN and Thirring models is very different, in
contradiction to results obtained with staggered
fermions~\cite{Chandrasekharan:2013aya,Chandrasekharan:2011mn}. The most obvious
distinction is that the GN model exhibits a phase transition at strong-coupling
to a phase in which a global symmetry (Z$_2$ in the example studied) is
spontaneously broken and a fermion mass dynamically generated; no such
transition is observed in the Thirring model despite strong evidence that the
physical strong coupling limit is probed. In the GN case, subsidiary
measurements of the axial Ward identity and the mass spectrum yielded results
consistent with large-$N$ expectations. This success may be due in part to the 
fields $\sigma$ and $\pi$ being related via equations of motion to bilinears
of direct interest such as the order parameter field; it is certainly the case that sampling the $\{\pi\}$
ensemble is a very effective means of estimating correlators formed from
disconnected diagrams. To our knowledge the measurement of the fermion
correlator is the first using DWF; 
compared to what is known using staggered
fermions, 
the contributions of excited state
artifacts are surprisingly large, 
necessitating an approach based on source smearing. This in turn
exacerbates the influence of phase fluctuations, so that the analysis needs to be
based on the quasi-Green functions $\tilde C$. It is remarkable that even so the
resulting spectrum shown in Fig.~\ref{fig:spectrum} matches large-$N$
expectations
as well as it does.

For the Thirring model, the non-observation of symmetry breaking is a robust
result suggesting that the critical number of flavors required for dynamical
symmetry breaking for the action (\ref{eq:LThir}) satisfies $N_c<2$. If we make
the additional assumption that the UV fixed point of the Thirring model
coincides with the IR fixed point of QED$_3$, then this is compatible with
the recent non-observation of a bilinear condensate in massless QED$_3$ with
$N\geq2$~\cite{Karthik:2015sgq}. 
Beyond this, by contrast, 
the picture is not satisfactory. Despite the
inapplicability of Elitzur's theorem, phase fluctuations
due to the use of smeared sources have precluded fermion spectroscopy, and
small fermion masses coupled with the absence of confinement have also prevented
success in meson channels. Neither failure invalidates the DWF approach; the
latter is simply a problem intrinsic to studying near-conformal physics on a
finite volume, while the former might in future  be tackled by a form of
gauge-fixing (see below). However, our inability to extract spectral information
severely curtails insight into the failure of the axial Ward
identity shown in Fig.~\ref{fig:ratio}, which potentially is a more profound
problem, though again not necessarily fatal. What is disappointing, though, is
that there is still no clear guide to the optimal lattice Thirring
formulation.

Obvious future directions to explore include different formulations of the
lattice Thirring model which may be less prone to the issues encountered here.
Ref.~\cite{Itoh:1994cr} highlighted the role of a ``hidden local symmetry''
which is manifest once the Thirring action (\ref{eq:LThirA}) is supplemented
by a scalar St\"uckelberg field $\phi$ coupled to $A_\mu$. The HLS model has a
gauge symmetry for which the Thirring model is the result of gauge-fixing to
$\phi=0$. Smoother gauge choices may enable better control over the fermion
propagator.  Formulating DWF with a true gauge symmetry will also restore
transversity of the vacuum polarisation, making identification of the
strong-coupling limit less ambiguous. Another route to finding and studying
critical behaviour might come via implemention
of the RHMC algorithm enabling $N=1$ to be simulated. However, a more promising
route, 
permitted by the control offered by the DWF
formulation, may be to introduce a U($2N$) and parity-invariant ``Haldane''
interaction term
$(\bar\psi\gamma_3\gamma_5\psi)^2$, motivated 
by the findings of the functional renormalisation
group~\cite{Gehring:2015vja} which identifies a significant Haldane component 
in the fixed-point action corresponding
to the Thirring model.  There is still much to learn about fermions in 2+1$d$.

\section*{Acknowledgements}
This work was supported by a Royal Society Leverhulme Trust Senior Research
Fellowship  LT140052, and in part by STFC grant ST/L000369/1. Numerical work was
performed on a PC cluster funded by Swansea University's College of Science. 
I have enjoyed discussions with Wes Armour, Ed Bennett,
Tony Kennedy and Tim Morris.

\appendix
\section{Free Fermion Propagator}
\label{app:A}

In this appendix we develop the propagator for the free DWF propagator in 2+1$d$
following the methods set out in~\cite{Shamir:1993zy} and \cite{Vranas:1997da}.
Inititally we consider the hermitian mass term $m_hS_h$.
In 2+1$d$ momentum space, the action may be written
\begin{equation}
S=\int_p\sum_{s,s^\prime}\bar\Psi(p,s)D(p;s,s^\prime)\Psi(p,s^\prime)
\end{equation}
where $D$ is related to the operator $D_0$ defined on a lattice with infinite
$s$-extent via
\begin{eqnarray}
D(p;s,s^\prime)&=&\theta(s-1)\theta(s^\prime-1)\theta(L_s-s)\theta(L_s-s^\prime)D_0(p;s,s^\prime)\nonumber\\
&+&m_h(P_+\delta_{s,1}\delta_{s^\prime,L_s}+P_-\delta_{s,L_s}\delta_{s^\prime,1}),
\end{eqnarray}
with 
\begin{eqnarray}
D_0&=&-(P_-\delta_{s+1,s^\prime}+P_+\delta_{s-1,s^\prime})+(b(p)+i{\bar p\!\!\!
/\,})\delta_{s,s^\prime};\nonumber\\
D^\dagger_0&=&-(P_+\delta_{s+1,s^\prime}+P_-\delta_{s-1,s^\prime})+(b(p)-i{\bar p\!\!\!
/\,})\delta_{s,s^\prime};\nonumber\\
(D_0D_0^\dagger)_{s,s^\prime}&=&\delta_{s,s^\prime}(1+b^2+\bar
p^2)-b(\delta_{s+1,s^\prime}+\delta_{s-1,s^\prime}),
\end{eqnarray}
and $\bar p_\mu$ and $b(p)$ defined in (\ref{eq:defs}).
The hermitian operator $D_0D_0^\dagger$ has zeromodes of the form
$\psi(s)=e^{\pm\alpha s}$:
\begin{equation}
D_0D_0^\dagger\psi(s)=[b^2(p)+\bar p^2-2b\cosh\alpha(p)+1]\psi(s),
\end{equation}
so the zero eigenvalue condition gives the definition of $\alpha$ in
(\ref{eq:defs}).
The Green function of $D_0D_0^\dagger$ is given by
\begin{equation}
G_0(s,s^\prime)={{e^{-\alpha\vert s-s^\prime\vert}}\over{2b\sinh\alpha}}\equiv
Be^{-\alpha\vert s-s^\prime\vert}.
\end{equation}

To find the Green function of $DD^\dagger$, we need to take into account both the
fact that the operators differ at $s=1$ and $s=L_s$, and the mass term coupling
the domain walls. Define
\begin{equation}
DD^\dagger=P_+\Omega_++P_-\Omega_-;\;\;G=P_+
G_++P_-G_-;
\end{equation}
then it can be verified that 
\begin{eqnarray}
\Omega_+(s,s^{\prime\prime})G_0(s^{\prime\prime},s^\prime)-\delta_{s,s^\prime}&=&
Be^{-\alpha
s^\prime}e^\alpha(be^{-\alpha}-1+m_h^2)\delta_{s,1}\nonumber\\
&+&Bbe^{-\alpha(L_s+1)}e^{\alpha
s^\prime}\delta_{s,L_s}\label{eq:missingh}\\
&+&Bm_hb[\delta_{s,1}e^{-\alpha(L_s-s^\prime)}+\delta_{s,L_s}e^\alpha e^{-\alpha
s^\prime}].\nonumber
\end{eqnarray}
and
\begin{eqnarray}
\Omega_+(s,s^\prime)e^{\pm\alpha s^\prime}&=&
\delta_{s,1}e^{\pm\alpha}(be^{\mp\alpha}-1+m_h^2)+m_hb\delta_{s,1}e^{\pm\alpha
L_s})\nonumber\\
&+&\delta_{s,L_s}(be^{\pm\alpha(L_s+1)}+m_hbe^{\pm\alpha}).\label{eq:extrah}
\end{eqnarray}
The $-$ conditions are obtained using the manifest symmetry
\begin{equation}
\Omega_-(s,s^\prime)=\Omega_+(L_s-s+1,L_s-s^\prime+1). 
\label{eq:symmetry}
\end{equation}
The general form of the propagator consistent with (\ref{eq:symmetry}) is then
\begin{eqnarray}
G_+(s,s^\prime)=G_0(s,s^\prime)&+&A_+e^{-\alpha(s+s^\prime-2)}+A_-e^{-\alpha(2L_s-s-s^\prime)}
\nonumber\\
&+&A_m(e^{_-\alpha(L_s+s-s^\prime-1)}+e^{-\alpha(L_s-s+s^\prime-1)}).
\label{eq:G+}
\end{eqnarray}
By requiring consistency for terms of the form $\delta_{s,1}e^{\mp s^\prime}$,
$\delta_{s,L_s}e^{\mp s^\prime}$, 
the condition\hfill\break
$\Omega_+(s,s^{\prime\prime})G_+(s^{\prime\prime},s^\prime)=\delta_{s,s^\prime}$
then yields the following equations:
\begin{equation}
{\cal C}
\begin{pmatrix}
{A_{+} \cr A_{m}}
\end{pmatrix}
=
B\left(\begin{matrix}{1-be^{-\alpha}-m_h^2\cr -m_hb\cr}\end{matrix}\right);\;\;
{\cal C}\left(\begin{matrix}{A_m\cr A_-\cr}\end{matrix}\right)=
B\left(\begin{matrix}{-m_hb\cr -be^{-\alpha}\cr}\end{matrix}\right)
\end{equation}
with
\begin{equation}
{\cal C}(m_h,L_s)=\begin{pmatrix}{
(be^\alpha-1+m_h^2)+m_hbe^{-\alpha(L_s-1)} & m_hb
+(be^{-\alpha}-1+m_h^2)e^{-\alpha(L_s-1)} \cr
m_hb +be^{-\alpha}e^{-\alpha(L_s-1)} & be^\alpha+m_hbe^{-\alpha(L_s-1)} \cr
}\end{pmatrix}.
\end{equation}
The solution is~\cite{Vranas:1997da}
\begin{eqnarray}
A_+&=&\Delta^{-1}B(e^\alpha-b)(1-m_h^2) \label{eq:A+}\\
A_-&=&\Delta^{-1}B(e^{-\alpha}-b)(1-m_h^2) \label{eq:A-}\\
A_m&=&\Delta^{-1}B[-2m_hb\sinh\alpha+e^{-\alpha(L_s-1)}(e^{-2\alpha}(b-e^\alpha)+m_h^2(e^{-\alpha}-b))]
\label{eq:Am}
\end{eqnarray}
with
\begin{eqnarray}
\Delta=b^{-1}\mbox{det}{\cal C}&=&
[e^{2\alpha}(b-e^{-\alpha})+m_h^2(e^\alpha-b)]\nonumber\\
&+&e^{-\alpha(L_s-1)}4m_hb\sinh\alpha\\
&+&e^{-2\alpha(L_s-1)}[m_h^2(b-e^{-\alpha})+e^{-2\alpha}(e^\alpha-b)].\nonumber
\end{eqnarray}

Next we explore the consequences of the anti-hermitian parity-invariant mass
term $m_sS_3$ (\ref{eq:m3S3})
so that now
\begin{eqnarray}
D(s,s^\prime)&=&\theta(s-1)\theta(s^\prime-1)\theta(L_s-s)\theta(L_s-s^\prime)D_0(s,s^\prime)\nonumber\\
&+&im_3P_+\delta_{s,1}\delta_{s^\prime,L_s}-im_3P_-\delta_{s,L_s}\delta_{s^\prime,1}.
\end{eqnarray}
Eqns. (\ref{eq:missingh},\ref{eq:extrah}) are replaced by
\begin{eqnarray}
\Omega^3_+(s,s^{\prime\prime})G_0(s^{\prime\prime},s^\prime)-\delta_{s,s^\prime}&=&
Be^{-\alpha
s^\prime}e^\alpha(be^{-\alpha}-1+m_3^2)\delta_{s,1}\nonumber\\
&+&Bbe^{-\alpha(L_s+1)}e^{\alpha
s^\prime}\delta_{s,L_s}\label{eq:missing3}\\
&+&iBm_3b[\delta_{s,1}e^{-\alpha(L_s-s^\prime)}-\delta_{s,L_s}e^\alpha e^{-\alpha
s^\prime}]\nonumber
\end{eqnarray}
and
\begin{eqnarray}
\Omega^3_+(s,s^\prime)e^{\pm\alpha s^\prime}&=&
\delta_{s,1}e^{\pm\alpha}(be^{\mp\alpha}-1+m_3^2)+im_3b\delta_{s,1}e^{\pm\alpha
L_s})\nonumber\\
&+&\delta_{s,L_s}(be^{\pm\alpha(L_s+1)}-im_3be^{\pm\alpha}),\label{eq:extra3}
\end{eqnarray}
while the symmetry (\ref{eq:symmetry}) is now
\begin{equation}
\Omega_-(s,s^\prime)=\Omega^*_+(L_s-s+1,L_s-s^\prime+1). 
\label{eq:symmetry3}
\end{equation}
motivating the {\it Ansatz}
\begin{eqnarray}
G_{3+}(s,s^\prime)=G_0(s,s^\prime)&+&A_+e^{-\alpha(s+s^\prime-2)}+A_-e^{-\alpha(2L_s-s-s^\prime)}
\nonumber\\
&+&A_{m_3} e^{_-\alpha(L_s+s-s^\prime-1)}+A_{m_3}^* e^{-\alpha(L_s-s+s^\prime-1)}.
\end{eqnarray}
The consistency conditions become
\begin{equation}
{\cal C}\left(\begin{matrix}{A_+\cr A_{m_3}^*\cr}\end{matrix}\right)=
B\left(\begin{matrix}{1-be^{-\alpha}-m_3^2\cr im_3b\cr}\end{matrix}\right);\;\;
{\cal C}\left(\begin{matrix}{A_{m_3}\cr A_-\cr}\end{matrix}\right)=
B\left(\begin{matrix}{-im_3b\cr -be^{-\alpha}\cr}\end{matrix}\right)
\end{equation}
with
\begin{equation}
{\cal C}(m_3,L_s)=\left(\begin{matrix}{
(be^\alpha-1+m_3^2)+im_3be^{-\alpha(L_s-1)} & im_3b
+(be^{-\alpha}-1+m_3^2)e^{-\alpha(L_s-1)} \cr
-im_3b +be^{-\alpha}e^{-\alpha(L_s-1)} & be^\alpha-im_3be^{-\alpha(L_s-1)} \cr
}\end{matrix}\right).
\end{equation}
The solutions (\ref{eq:A+}), (\ref{eq:A-}) remain valid with $m\leftrightarrow
m_3$ and $\Delta\leftrightarrow\Delta_3$, but now
\begin{equation}
A_{m_3}=\Delta_3^{-1}B[-2im_3b\sinh\alpha+e^{-\alpha(L_s-1)}(e^{-2\alpha}(b-e^\alpha)+m_3^2(e^{-\alpha}-b))]
\label{eq:Am3}
\end{equation}
with
\begin{eqnarray}
\Delta_3=b^{-1}\mbox{det}{\cal C}&=&
[e^{2\alpha}(b-e^{-\alpha})+m_3^2(e^\alpha-b)]\label{eq:Delta3}\\
&+&e^{-2\alpha(L_s-1)}[m_3^2(b-e^{-\alpha})+e^{-2\alpha}(e^\alpha-b)].\nonumber
\end{eqnarray}
Two features are apparent: first, there are no  O$(e^{-\alpha L_s})$
contributions to $\Delta_3$, so the first correction is O$(e^{-2\alpha L_s})$; 
second, the O$(e^{-\alpha L_s})$ contribution to
$A_{m_3}$ is now shifted by a phase $e^{i{\pi\over2}}$ with respect to the leading
order piece. Both features mitigate finite-$L_s$
corrections to the calculation of $\langle\bar\psi\gamma_3\psi\rangle$ in the
large-$N$ GN model presented in Sec.~\ref{sec:1/N}.

\end{document}